\documentclass[preprint2]{aastex6}
\usepackage{amsmath}
\usepackage{bm}
\usepackage{graphicx}
\usepackage{amssymb}
\usepackage{gensymb}
\usepackage{array}
\usepackage{color}

\begin{document}
  
\title{Saturn's formation and early evolution at the origin of Jupiter's massive moons}

\author{T. Ronnet, O. Mousis, P. Vernazza}
\affil{Aix Marseille Univ, CNRS, LAM, Laboratoire d'Astrophysique de Marseille, Marseille, France {\tt thomas.ronnet@lam.fr}}
 \author{J. I. Lunine}
 \affil{Department of Astronomy and Carl Sagan Institute, Cornell University, Ithaca, NY 14853, USA}
 \author{A. Crida}
 \affil{Universit{\'e} C{\^o}te d'Azur/Observatoire de la C{\^o}te d'Azur, Laboratoire Lagrange (UMR7293), Boulevard de l'Observatoire, CS 34229, 06300 Nice, France\\
 Institut Universitaire de France, 103 Boulevard Saint-Michel, 75005 Paris, France}
  
\begin{abstract}
The four massive Galilean satellites are believed to have formed within a circumplanetary disk during the last stages of Jupiter's formation. While the existence of a circum-jovian disk is supported by hydrodynamic simulations, no consensus exists regarding the origin and delivery mechanisms of the building blocks of the forming satellites. The opening of a gap in the circumsolar disk would have efficiently isolated Jupiter from the main sources of solid material. However, a reservoir of planetesimals should have existed at the outer edge of Jupiter's gap, where solids were trapped and accumulated over time. Here we show that the formation of Saturn's core within this reservoir, or its prompt inward migration, allows planetesimals to be redistributed from this reservoir towards Jupiter and the inner Solar System, thereby providing enough material to form the Galilean satellites and to populate the Main Belt with primitive asteroids. We find that the orbit of planetesimals captured within the circumjovian disk are circularized through friction with gas in a compact system comparable to the current radial extent of the Galilean satellites. The decisive role of Saturn in the delivery mechanism has strong implications for the occurrence of massive moons around extrasolar giant planets as they would preferentially form around planets within multiple planet systems.  
\end{abstract}

  \section{Introduction}
 All four giant planets of the Solar System possess regular satellites which likely formed in situ \citep{PC15}. The origin of the satellites systems can therefore provide hints on the conditions prevailing at the epoch of their formation. Whereas the regular satellites of Saturn, Uranus and Neptune may well have formed after their host planets, by the spreading of massive rings \citep{CC12}, the Galilean satellites are generally seen as a byproduct of Jupiter's formation \citep[e.g.,][]{LS82,Co95,CW02,ME03a,ME03b,MG04} and could therefore help in better understanding how, where and when the giant planet formed. Moreover, the inferred existence of a water ocean underneath the icy crust of Europa and likely within Ganymede and Callisto make the Galilean system of peculiar interest from an astrobiological point of view and motivate the search for potentially habitable extrasolar moons \citep[for a review see e.g.,][]{He14}. Yet, some crucial steps need to be unveiled to assess the origin of the Galilean satellites and the likelihood of finding similar objects around giant exoplanets.
    
In the current paradigm, the jovian satellites would have formed within a circumplanetary disk (CPD) that surrounded Jupiter at the very end of its formation \citep[see e.g.,][for a review]{CW09,EM09}. Although the development phase of a CPD and its precise structure are not well constrained, its existence around a Jupiter mass planet has been well established through numerical experiments \citep[e.g.,][]{Ma08,TOM12}. However, the fundamental issue of the origin of the solids embedded within the jovian CPD remains. Although several mechanisms of solids delivery have been proposed, no consensus currently exists. This is problematic because how solids are brought to the CPD in turn dictates their initial mass and size distributions which then essentially determine the accretion timescale of the satellites and their final masses.  
    
Formation models of the Galilean satellites generally fall in two distinct classes, the so-called gas starved model \citep[e.g.,][]{CW02} and the minimum mass subnebula model \citep[e.g.][]{LS82,ME03a,ME03b}, each being associated with a different source of solids and delivery mechanism. In the starved disk model of \citet{CW02}, Jupiter is still feeding from the circumsolar disk at the formation epoch of its satellites and its CPD is constantly replenished with fresh material. Canup \& Ward argue that small solids are entrained with the gas inflow onto the CPD and provide the bulk material necessary to form the satellites over a timespan of 10$^5$ to 10$^6$ years. Minimum mass models, on the other hand, are {\it ad hoc} constructions of a disk where sufficient condensable material to form the satellites is augmented with gas upon reaching a solar composition. Contrary to starved disk models, this yields very dense gaseous disks and a rapid assemblage of the satellites in $10^2$--$10^4$ years \citep{LS82}. \citet{ME03a,ME03b} revisited such models by enhancing the solid mass fraction by a factor 3--4 to account for the enrichment over solar abundances observed in Jupiter's atmosphere and adding an extended outer disk leading to a longer formation timescale of the satellites (especially Callisto). They argued that a dense CPD provides suitable conditions for the capture/ablation of planetesimals ($\gtrsim $ 10 m) on initially heliocentric orbits close to Jupiter which would have provided the bulk material necessary to form the satellites \citep[see also the discussion in][]{EM09}.
    
In both scenarios, the delivery of solids to the CPD is tightly linked to the formation history of Jupiter and the distribution of dust/planetesimals in its vicinity. In recent years, large strides have been made in the theory of planet formation. Of particular interest are numerous recent studies that have demonstrated the efficiency of the so-called pebble accretion, i.e. the gas drag assisted accretion of $\sim$cm sized solids, in growing the giant planets cores \citep{OK10,JL10,LJ12,LJ14,LKD15}. This new formation paradigm implies that most of the solid mass budget in the forming giant planet region was contained in pebbles and not in larger planetesimals. As the pebbles are very sensitive to aerodynamic drag, their distribution within the disk does not necessarily follow a power-law distribution, such as that advocated in the widely used minimum mass solar nebula model \citep[e.g.][]{Ha81}, and is affected by pressure perturbations. This issue is crucial for understanding the origin of the jovian massive moons as the spatial and size distributions of the solids in the vicinity of Jupiter set the conditions of their delivery to the CPD.
   
Moreover, it is now established that the growth and dynamics of the giant planets have a tremendous influence on the distribution of small bodies within the Solar System \citep[see e.g.,][]{Go05,Mo05,Le09,Wa11,VBN16,RI17}. Despite this fact, the formation of the jovian moons in the broader context of the early history of the giant planets in the protoplanetary disk has not been quantitatively investigated. 

These considerations motivated the present study, in which we attempt to address the delivery of solid material to the circumjovian disk in light of the recent theories of giant planet formation \citep[e.g.,][]{OK10,LJ14,LJM14,LKD15}. In Section~\ref{context}, we discuss some of the limitations of the proposed delivery mechanisms and introduce our framework. We thus propose that the building blocks of the Galilean satellites originated from a reservoir of planetesimals located at the outer edge of the gap opened by Jupiter in the circumsolar disk. However, because this reservoir remains mainly out of Jupiter's reach, we show in Section~\ref{results} the decisive role of Saturn's growth and early evolution. The forming Saturn, we show, had the potential to perturb the planetesimals' orbits and to allow their delivery to both the jovian CPD and the inner Solar System. In Section~\ref{CPD}, we investigate the subsequent evolution of the planetesimals within the circumjovian disk. The implications of our results along with some additional considerations raised by the model are discussed in Section~\ref{discussion} and our findings are summarized in Section~\ref{summary}.

\section{Sources of solid material}\label{context}
    
Here we briefly present the prevailing scenarios for the origin of the building blocks of the Galilean satellites and discuss some of their limitations. Considering the hurdles of the proposed mechanisms, we argue for the existence of a reservoir of material located at the outer edge of Jupiter's gap. This reservoir likely provided the bulk of the material for the Galilean satellites as will be shown in the next sections.
        
\subsection{Inflow of small dust grains} 
\label{CW}
    
\citet{CW02} postulated that the Galilean satellites formed while Jupiter was still feeding from the circumsolar disk via the replenishment of its CPD with a mixture of gas and dust in solar proportions. This model was originally proposed to circumvent some weak points of the minimum mass models, specifically the long accretion timescale needed to match the internal structure of Callisto and the survival of satellites against gas-driven migration. 
    
However, the scenario of \citet{CW02} requires that the solids brought to the CPD were in the form of perfectly coupled dust grains that have not settled towards the midplane of the disk. Indeed, hydrodynamic simulations demonstrated that the gas eventually falling onto the CPD resides well above the midplane of the circumsolar disj \citep{Ma08,TOM12,Sz14,Mo14}. The dust grains that substantially grew up and settled towards the midplane of the disk due to gas drag would therefore not be able to reach the CPD with the characteristics defined by \citet{CW02}. \citet{PM06} and \citet{Pa07} have shown that only particles with sizes $\leqslant$10 $\mu$m could be entrained with the gas flow once Jupiter opened up a gap in the circumsolar disk. \citet{BOD11,BKE12} precisely investigated dust growth within protoplanetary disks and found that it is efficient at least up to partially decoupled sizes (mm to cm, depending on the turbulence level and location in the disk), implying a substantial settling of dust grains towards the disk's midplane. Considering the results of \citet{BOD11}, \citet{Zh12} estimated the dust-to-gas ratio within the gap opened by a Jupiter mass planet to be 10$^{-4}$, which is two orders of magnitude lower than the protosolar value. On the other hand, \citet{Sh17} studied grain growth within CPDs and find that the dust-to-gas ratio must be $\gtrsim 1$ in order to grow satellitesimals via direct collision or streaming instability \citep[i.e., the collapse of a cloud of pebbles concentrated through gas drag into $\sim$100 km objects,][]{Jo15}. 

Considered together, it is difficult to reconcile these results with the scenario envisioned by \citet{CW02}. The gas accreted by Jupiter and the CPD was most likely depleted in dust and might not have provided the bulk of the material necessary to form the satellites. 
      
\subsection{Capture of large planetesimals} \label{ME}
      
Another potential mechanism to deliver solid material to the CPD is the capture/ablation of larger planetesimals located in the vicinity of Jupiter due to either collisions in a gas poor environment \citep{EM06} or gas drag within a gas rich CPD \citep{MEC10}.
The latter process has been numerically investigated by several authors \citep{Fu13,SOF16,SO17,DaP15}.
However, the existence of planetesimals in the close vicinity of Jupiter is questionable. It is now well known that a planet as massive as Jupiter should have carved a deep gap in the circumsolar disk \citep[e.g.,][]{LP86}. The opening of a gap in the planetesimal or dust distribution will predate the opening of a deep gap in the gas distribution \citep[e.g.,][]{LTD10,Pa07,LJM14,Di16,DL17}. Unless subject to a replenishment mechanism, the feeding zone of Jupiter should have been rapidly devoid of solid material. As a matter of fact, \citet{SO17} pointed out that if a gap existed in the planetesimal distribution beyond the orbit of Jupiter, the accretion of material onto the CPD would be greatly reduced, if not supressed. This is a crucial issue considering that the Galilean satellites should have formed in the later stages of Jupiter's formation \citep[e.g.,][]{CW09,EM09}.
      
The existence of a sea of planetesimals in the giant planet region to feed Jupiter's disk also remains hypothetical. In the current paradigm of planetesimal formation \citep[see e.g.,][]{Jo14}, specific conditions need to be fulfilled for large bodies to form, resulting in potentially very localized regions of efficient planetesimal formation \citep{DAM16,Ca17,ScO17}. At first sight, it seems that the opening of a gap by Jupiter is problematic for the formation of its satellites as this would have substantially isolated the giant planet from any source of solid material.
      
\subsection{Existence of a reservoir of planetesimals close to Jupiter} 
\label{reservoir}
      
Recent developments in the theory of giant planets formation suggest that the rapid formation of a solid core of several Earth masses is facilitated if the solid mass budget of protoplanetary disks (PPDs) is carried by dust grains only partially decoupled from gas, designated as pebbles, with sizes in the mm--cm range\footnote{We stress that the sizes given here are mere indications as pebbles are defined by their aerodynamic properties and not their sizes.} 
\citep{LJ12,LJ14,LKD15}. The very efficient accretion of pebbles leads to high mass accretion rates and substantial heating of the envelope that prevents its rapid contraction onto the core. Pebble accretion is however halted when the core becomes massive enough so that it perturbs significantly the surrounding gas distribution, creating a pressure maximum outside its orbit that acts as a barrier \citep{LJM14}. After reaching this mass threshold, the accretional heating of the core's envelope ceases, allowing a rapid contraction of the atmosphere of the protoplanet and its subsequent growth toward becoming a gas giant.
     
Once Jupiter reached the pebble isolation mass \citep[estimated to be $\sim$20$\;M_\oplus$;][]{LJM14}, pebbles remained trapped at the outer edge of its gap and accumulated over time \citep[e.g.,][]{Go15}. The accumulation of solids at this particular place would have lead to an enhanced dust-to-gas ratio and therefore likely provided suitable conditions to trigger the formation of large planetesimals via direct sticking or gravitational instability. Therefore, a reservoir of planetesimals should have built up over time just outside Jupiter's orbit, while the close vicinity of the planet was devoid of solid material. This reservoir is potentially so massive that \citet{KOI12} proposed that Saturn's core actually grew at the outer edge of Jupiter's gap \citep[an hypothesis also mentioned by][]{LJM14}.
     
It would be surprising were such a reservoir to have existed close to Jupiter and not play any role in the formation of its regular satellites, the origin of whose building blocks remains elusive. Yet, as demonstrated by \citet{SO17}, if the objects of the reservoir were on circular and coplanar orbits, as expected from their formation process, they would have mainly remained out of Jupiter's reach. However, there is now little doubt that Saturn once was orbiting much closer to Jupiter than it is currently \citep[see e.g.,][and references therein]{De17}. Saturn could therefore have had a great influence on the dynamics of the planetesimals residing at the outer edge of Jupiter's gap, exciting their orbits and potentially allowing their delivery to the jovian CPD. This idea constitutes the cornerstone of the present study.

\begin{figure}
\includegraphics[width=\linewidth]{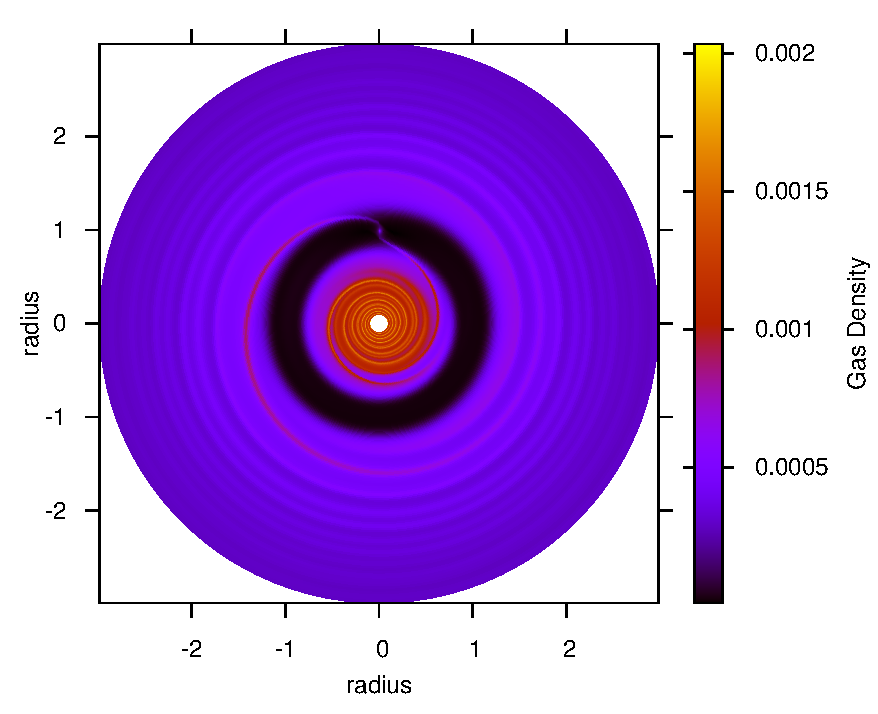}
\caption{Fargo simulation of a Jupiter mass planet in a viscous disk with a constant aspect ratio of 0.05 (i.e., the scale height of the disk normalized by the orbital distance). The turbulent viscosity was accounted for following the prescription of \citet{SS73} with $\alpha=2\times10^{-3}$. The radius is expressed in terms of the giant planet's semi-major axis and the gas density is in arbitrary units. This gas distribution is obtained after 300 orbits of the planet.}
\label{FARGO}
\end{figure}

\section{Delivering planetesimals from the reservoir} \label{results}
  
Here we investigate the orbital evolution of the planetesimals trapped at the outer edge of Jupiter's gap and under the influence of both the planet itself (assumed to have acquired essentially its current mass) and the forming Saturn.

Jupiter is assumed to be located at a heliocentric distance of $\sim$5.4 AU, in agreement with the dynamical evolution of the giant planets after the dispersal of the circumsolar disk \citep[][and references therein]{De17}. This does not imply that Jupiter never suffered from any migration within the disk. Rather, the planet migration rate was substantially lowered when it opened up a gap in the disk \citep{LP86,CB17}, so that the reservoir of planetesimals could have built up over time.
  
Regarding Saturn, we explored two different evolution pathways, first because of the many unknowns of its formation history and second, to show that the redistribution of solids from the reservoir is a natural outcome and does not necessarily require very specific configurations of Jupiter and Saturn. In Section~\ref{gap}, we investigate a scenario where the core of Saturn is formed at the outer edge of Jupiter's gap, as proposed by \citet{KOI12}. Alternatively, Saturn could have formed further from Jupiter and migrated inward until being caught in resonance with Jupiter \citep[e.g.,][]{BLJ15}. We explore this possibility in Section~\ref{migration}.
  
The orbital integrations were performed using the hybrid {\tt HERMES} integrator available with the open source {\tt REBOUND} package\footnote{Available at http://github.com/hannorein/rebound}. Each simulation included 5,000 planetesimals as test particles and the orbits were integrated with a timestep of 10$^{-2}/2\pi$ yr\footnote{ We note that this is the timestep for the simplectic integrator only. Close encounters with the massive planets are handled with the high order adaptive timesteping {\tt IAS15} integrator \citep{RT15,RS15}}. In each case, we included the eccentricity damping of the giant planets due to interaction with the gas disk using fictitious forces (Appendix~\ref{extra}). We used disk profiles including a Jupiter mass planet and associated gap obtained from 2D hydrodynamic simulations performed with FARGO \citep{Ma00}. Figure~\ref{FARGO} shows the gas distribution obtained after 300 orbits of Jupiter. We normalized the disk profiles so that the surface density at 1 AU is $\sim$300 $\mathrm{g\,cm^{-2}}$, which corresponds to a moderately evolved disk \citep{Bi15}. We included the effect of aerodynamic drag in the equation of motion of the planetesimals, considering they have a radius of 100 km and a density of 1 $\mathrm{g\,cm^{-3}}$(see Appendix~\ref{extra}). When planetesimals were found at a distance $r \leqslant 150 \, R_\mathrm{Jup}$ from Jupiter, the aerodynamic drag was computed with respect to a CPD profile derived from the parameterization of \citet{Sa10}. A description of the CPD model is provided in Appendix~\ref{CPDmod}.
  
We consider that planetesimals are captured within the circum-jovian disk when they are found on a bound orbit with a semi-major axis with respect to Jupiter that is less than 0.2 $R_\mathrm{Hill}$, where $R_\mathrm{Hill} = a_\mathrm{Jup}(M_\mathrm{Jup}/{3M_\odot})^{1/3}$ is the Hill's radius of Jupiter. This quite arbitrary threshold was chosen because it corresponds roughly to the extension of the circum-jovian disk and is plausibly deep enough in Jupiter's potential to consider the objects as permanently captured within the CPD. More details about the capture of planetesimals and a test of the validity of the threshold are presented in Appendix~\ref{capture}. The orbital parameters of the planetesimals with respect to the Sun or Jupiter are computed using the dedicated tools provided in the {\tt REBOUND} package. Captured planetesimals are removed from the simulation to save computing power and their orbital parameters with respect to Jupiter are stored.
  
\subsection{Case 1: growth of Saturn at the edge of Jupiter's gap}\label{gap}
  
\begin{figure*}
\includegraphics[width=\linewidth]{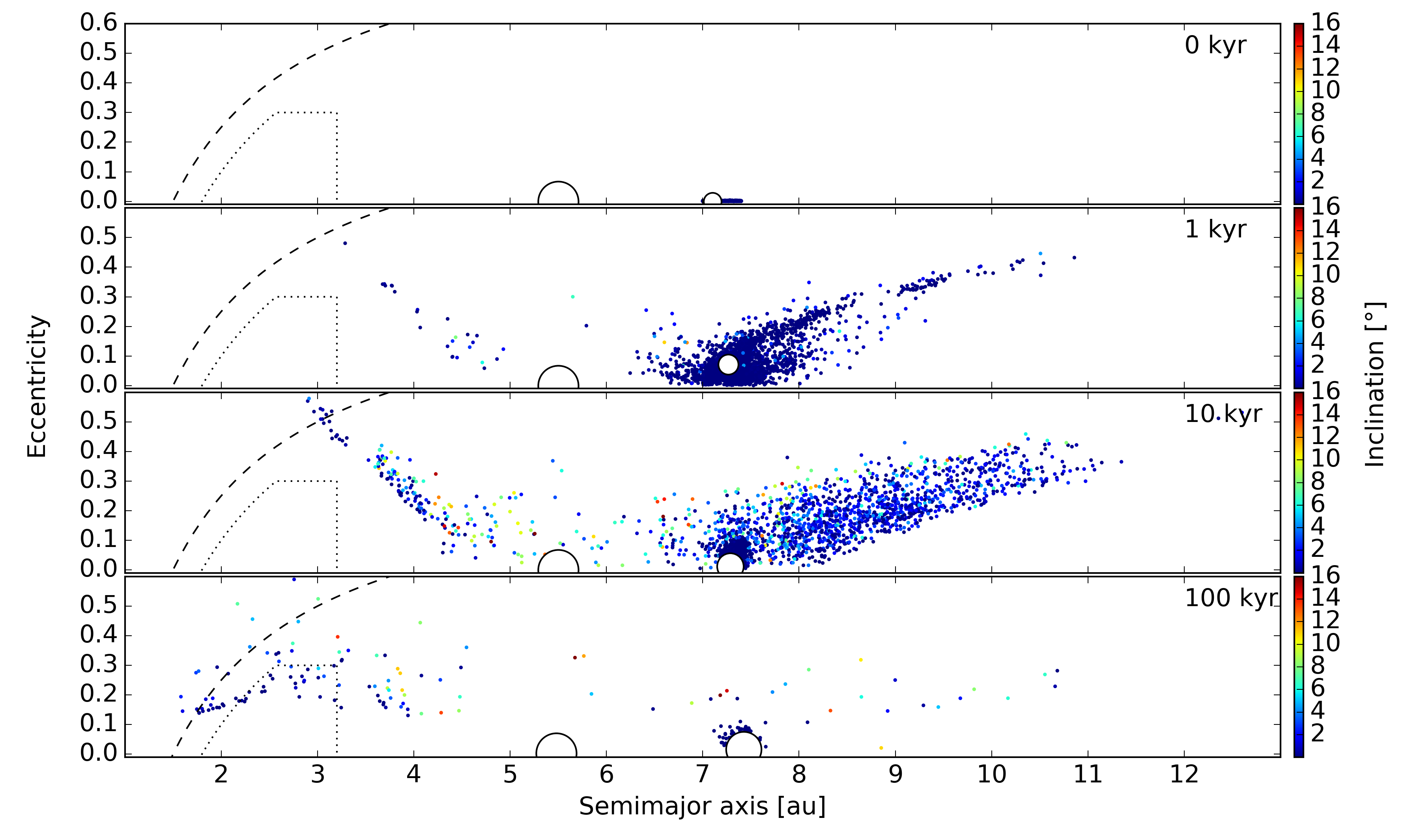}
\caption{Orbital evolution of the planetesimals with Saturn growing at the outer edge of the gap over a timescale $\tau_\mathrm{growth}=5\times10^5$ yr. The orbits of the planetesimals, initially nearly circular, are excited by the growing planet and scattered both inwards and outwards. The excitation of the eccentricity of the planetesimals allows their capture within the circumjovian disk and injection in the inner Solar System. The dotted box roughly represents the extension of the asteroid belt while the dashed line marks the orbits with $q = 1.5$ AU. Planetesimals with a perihelion $q \leqslant 1.5$ AU would interact with the embryos of the terrestrial planets, and potentially deliver water to them.}
\label{case1}
\end{figure*}
  
Here we present the results of simulations considering the growth of a body from a mass of $\sim$1 M$_\oplus$ up to the mass of Saturn and located at a heliocentric distance of 7 AU (with Jupiter placed at 5.4 AU). The mass of the protoplanet, $M_\mathrm{Sat}$, is increased on a timescale $\tau_\mathrm{growth}$ ranging between $10^5$--$10^6$ years following:
  
\begin{equation}\label{Mdot}
M_\mathrm{Sat}(t) = M_i + \Delta M\left[1 - \exp(-t/\tau_\mathrm{growth})\right],
\end{equation}

\noindent where $M_i$ is the initial mass of the core and $\Delta M$ is the difference between the initial core mass and the final mass of Saturn.
This evolution pathway is very simplified compared to the core accretion model where an envelope is slowly contracted until a rapid runaway gas accretion is triggered and then followed by a slower accretion phase when the planet carves a gap in the disk \citep[e.g.,][]{Po96}. However, the classical picture of core accretion might be inaccurate due to the fact that the gas and solids distributions are significantly perturbed in the particular case considered here. Detailed investigations would be needed to obtain a more realistic growth pattern but we do not aim here at studying the precise evolution of Saturn. We nevertheless varied the growth timescale to see whether some trends stand out in the final planetesimals distribution.
  
Figure~\ref{case1} shows the orbital evolution in the semimajor axis-eccentricity plane obtained from a simulation with Saturn growing over a timescale $\tau_\mathrm{growth}$ = $5\times10^5$ years. The eccentricity of the planetesimals is excited by Jupiter and the growing core, allowing them to cross Jupiter's orbit and be redistributed inwards or outwards. Some of the planetesimals are implanted in the main asteroid belt, whose boundaries are illustrated by the dotted box in Figure~\ref{case1}, and others have orbits that cross the region of terrestrial planets embryos (which were not included in the simulation) marked by the dashed line (see figure legend for details). Issues regarding the implantation of objects in the inner Solar System are further discussed in Section~\ref{discussion}. Here, we are more concerned with the capture of planetesimals within the circumjovian disk.
  
A matter of critical importance is the relative number of objects captured by Jupiter with respect to that of objects implanted in the Main Belt. Currently, the mass of the asteroid belt is estimated to be $\sim$$5\times10^{-4}\;$M$_\oplus$ \citep{Kr02} whereas the mass of the Galilean system is approximately $\sim$$6\times10^{-2}\;$M$_\oplus$. Although it is expected that the asteroid belt has been depleted in mass throughout its history \citep{Mo15}, a scenario where more mass is implanted in the asteroid belt than in the CPD would be hardly reconcilable with the two orders of magnitude more massive Galilean system observed today. Moreover, it is very likely that the accretion of the jovian moons was far from being perfectly efficient, implying that more than the current mass of the Galilean system should have been embedded within the CPD.
  
The results of the simulations with different growth timescales are summarized in Table~\ref{eff1}. The CPD capture and Main Belt implantation efficiencies are expressed as a percentage of the total number of objects initially located at the outer edge of Jupiter's gap. In all the cases investigated, we find that approximately one order of magnitude more objects end up captured within the CPD rather than being implanted in the Main Belt. We also note that some planetesimals directly collide with Jupiter in our simulations and would be subsequently ablated in its envelope, in proportions similar to that of the captured objects. The higher capture efficiency was obtained for Saturn growing on a 5$\times10^5$ years timescale.
In this case, considering that a mass equivalent to that of the Galilean system ($\sim$$6\times10^{-2}\,M_\oplus$) was captured by Jupiter implies an initial mass of planetesimals of $\sim$0.41$\,M_\oplus$ in the reservoir and $\sim$5.3$\times10^{-3}\,M_\oplus$ of material implanted in the main asteroid belt. Considering the efficiencies obtained from different growth timescales yield very similar results with an initial reservoir mass varying from $\sim$0.41 to 0.69$\,M_\oplus$ and a mass implanted in the asteroid belt varying from $\sim$5.3$\times10^{-3}$ to $\sim$9.2$\times10^{-3}\,M_\oplus$. These values are crude order of magnitude estimates as the mass captured within the CPD should be higher than that of the Galilean satellites, unless the accretion was perfectly efficient.
The mass implanted in the asteroid belt nevertheless compares well with that estimated in the Grand Tack scenario of \citet{Wa11}. These authors find a final asteroid belt containing $\sim$4$\times10^{-3}\,M_\oplus$ of planetesimals originating from beyond Jupiter's orbit. 
  
\begin{deluxetable}{ccc}
\tabletypesize{\scriptsize}
\tablecaption{CPD capture and Main Belt implantation efficiencies for Case 1 scenario}
\tablehead{
\colhead{$\tau_\mathrm{growth}$ (yr)} & \colhead{Capture} & \colhead{Implantation}
}
\startdata
 $1\times10^5$ & 8.7\% & 0.9\% \\
 {} & {} \\
 $5\times10^5$ & 14.8\% & 1.3\% \\
 {} & {} \\
 $1\times10^6$ & 11.8\% & 1.8\%\\
\enddata
\label{eff1}
\end{deluxetable}

\subsection{Case 2: migration of Saturn towards Jupiter}\label{migration}
  
\begin{figure*}
\includegraphics[width=\linewidth]{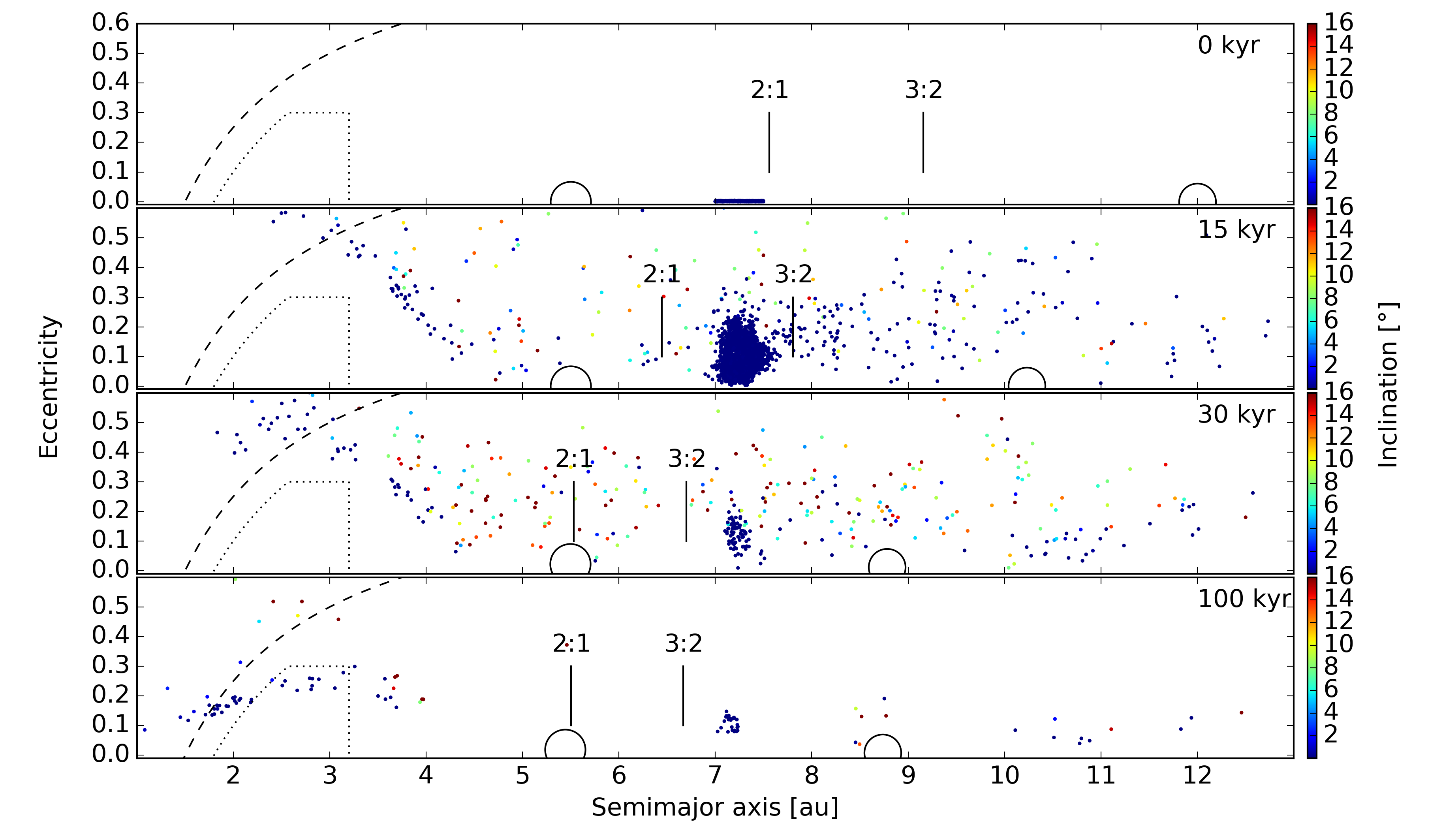}
\caption{Orbital evolution of planetesimals with Saturn migrating towards Jupiter over a timescale $\tau_\mathrm{mig}= 10^5$ years. The small vertical lines, labelled 2:1 and 3:2, show the positions of the corresponding MMRs with Saturn. The dashed line and the dotted box are equivalent to those of Figure~\ref{case1}. The planetesimals are excited when the reservoir is swept out by the 2:1 and 3:2 MMRs with Saturn after 15 and 30 kyr, respectively.}
\label{case2}
\end{figure*}
  
Another plausible scenario is that Saturn formed further from Jupiter and migrated inwards rapidly (before possibly opening its own gap), thereby catching up with Jupiter until the giants were caught in a mean motion resonance (MMR). 
Contrary to Case 1, this scenario does not constrain a precise location for the formation of Saturn. The formation of Saturn in the more distant regions of the disk could be the mere result of the initial distribution of material in the disk and the stochastic nature of accretion \citep[e.g.,][]{LKD15}, or, it could be the result of self-organization in the disk when Hall effect is considered. The self-organization results in zonal flows which naturally creates axisymmetric dust traps at different radial distances whose number and locations depends on the magnetic flux and intensity of the Hall effect \citep{BLF16}.

To investigate such a scenario, we conducted simulations where Saturn started at 12 AU and then migrated on different timescales towards Jupiter. We considered a fully formed Saturn to highlight the effect of the migration timescale on the final distribution of planetesimals. We mimicked the migration of Saturn by applying a fictitious force acting on a timescale $\tau_\mathrm{mig}$, which yields the following acceleration term \citep[e.g.,][]{CN08}:

\begin{equation}
{\bf a}_\mathrm{mig} = -\frac{\bf v}{\tau_\mathrm{mig}}.
\end{equation}

\noindent For the sake of simplicity, we turned off the force when Saturn is caught in the 2:1 MMR with Jupiter to avoid unphysical crossing of the resonance. 
Whether Jupiter and Saturn end up in their mutual 2:1 or 3:2 MMR is nevertheless not critical for the delivery of planetesimals, as shown below. Also, given the many uncertainties in the formation history of the giant planets and considering our very simplified model, we do not aim here at exploring the full range of possible parameters. 
  
Figure~\ref{case2} shows snapshots of the evolution of the system with Saturn migrating on a timescale $\tau_\mathrm{mig} = 10^5$ years. The sweeping of the reservoir of planetesimals by the 2:1 and 3:2 MMRs with Saturn excites the planetesimals' orbits and allows their delivery to the jovian CPD and the inner Solar System. The vast majority of planetesimals have been redistributed after the passage of the 3:2 MMR with Saturn across the reservoir.
  
The percentage of objects captured within the CPD and implanted in the main asteroid belt at the end of the simulations for different migration timescales of Saturn are summarized in Table~\ref{eff2}. The capture efficiencies differ from case to case due to the fact that the excitation of the eccentricity of the planetesimals in MMR with Saturn depends on the velocity of the giant planet. In the case where Saturn migrates on a $5\times10^5$ years timescale, the planetesimals are efficiently captured in the 2:1 MMR and reach very high eccentricities.
  
In the other cases, the planetesimals are only excited by the 2:1 MMR, they are not captured, and reach lower eccentricities. Therefore, more objects with lower eccentricities remain when the 3:2 MMR with Saturn sweeps the reservoir and this yields slightly higher capture efficiencies. Nevertheless, the differences are not dramatic. The percentage of captured objects varies from $\sim$14.4\% in the most favorable case down to $\sim$9\% for the slow migration case, assessing the robustness of the mechanism against the range of plausible migration rates of Saturn. The implantation of objects in the Main Belt is also comparable for each investigated migration rate with efficiencies that are more than one order of magnitude lower than the CPD capture efficiencies. Similarly to the Case 1 scenario, we find that a number of objects equivalent to that of the captured planetesimals directly collide with Jupiter.
  
We find that a migration rate $\tau_\mathrm{mig}= 10^5$ years yields the highest capture efficiency within the circum-jovian disk with $\sim$14.4\% of planetesimals from the reservoir captured. Considering the captured objects represent the mass of the Galilean system ($\sim$6$\times10^{-2}\,M_\oplus$), the initial reservoir should have had a mass of $\sim$0.42 $M_\oplus$ and the mass implanted in the asteroid belt would be $\sim$1.7$\times10^{-3}\,M_\oplus$. With the different efficiencies derived, we find that the intial mass of the reservoir would vary from $\sim$0.42 to 0.67 $M_\oplus$ and the mass implanted in the asteroid belt from $\sim$1.7$\times10^{-3}$ to $3.3\times10^{-3} \, M_\oplus$. These results are very similar to those obtained in the Case 1 scenario with the notable difference that the implantation of objects in the asteroid belt is less efficient.

\begin{deluxetable}{ccc}
\tabletypesize{\scriptsize}
\tablecaption{CPD capture and Main Belt implantation efficiencies for Case 2 scenario}
\tablehead{
\colhead{$\tau_\mathrm{mig}$ (yr)} & \colhead{Capture} & \colhead{Implantation}
}
\startdata
 $5\times10^4$ & 12.9\% & 0.6\% \\
 {} & {} \\
 $1\times10^5$ & 14.4\% & 0.4\% \\
 {} & {} \\
 $5\times 10^5$ & 9.0\% & 0.5\%\\
\enddata
\label{eff2}
\end{deluxetable}

\section{Evolution of captured planetesimals} \label{CPD}
  
\begin{figure}
\includegraphics[width=\linewidth]{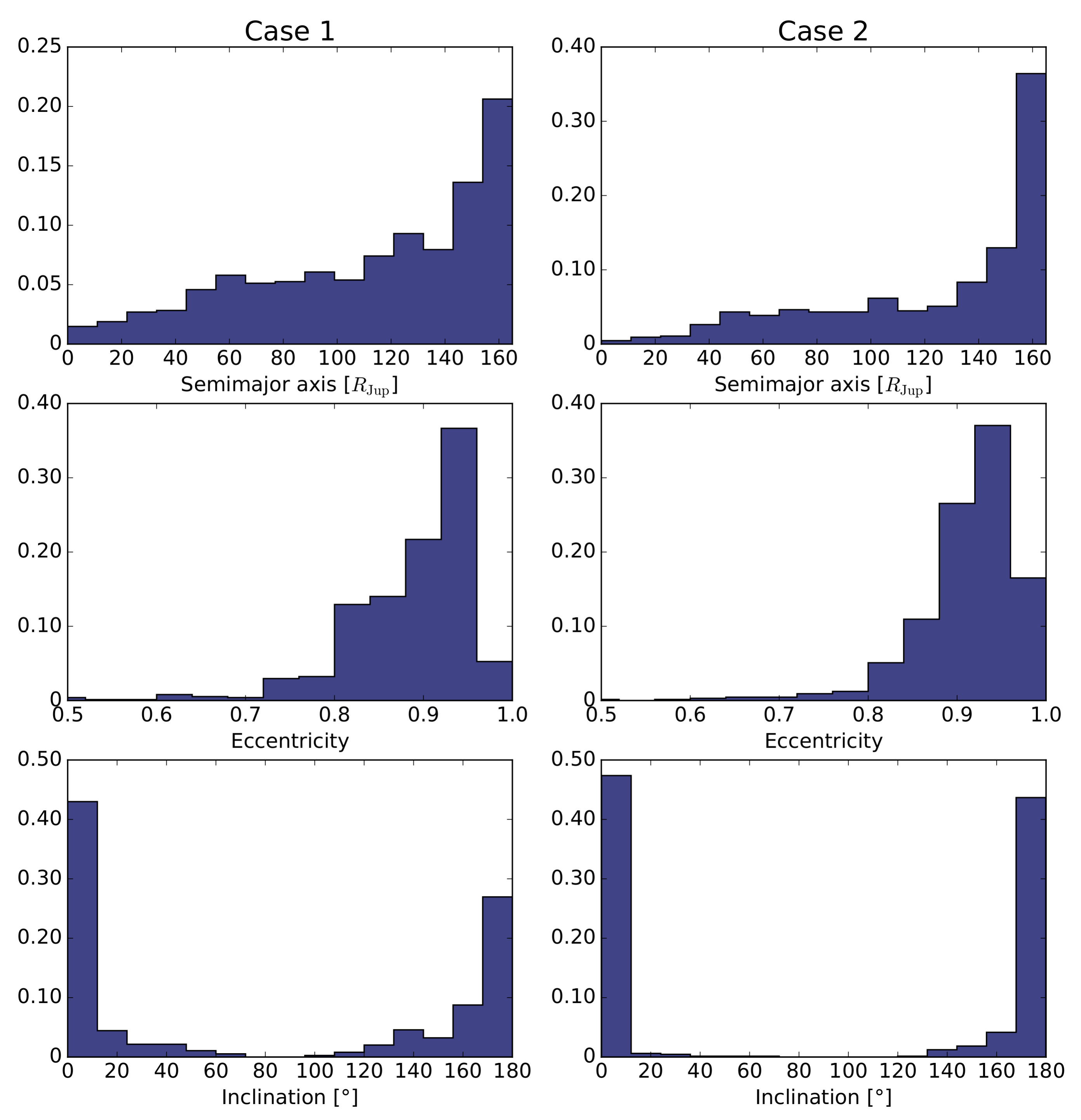}
\caption{Comparison of the orbital parameters of the captured objects in Case 1 with $\tau_\mathrm{growth}=5\times10^5$ years (left) and in Case 2 with $\tau_\mathrm{mig}=10^5$ years (right). The histograms are normalized according to the total number of captured objects. Both cases exhibit very similar trends with planetesimals initially captured on large, very eccentric and inclined orbits.}
\label{initial}
\end{figure}

\begin{figure}
\includegraphics[width=\linewidth]{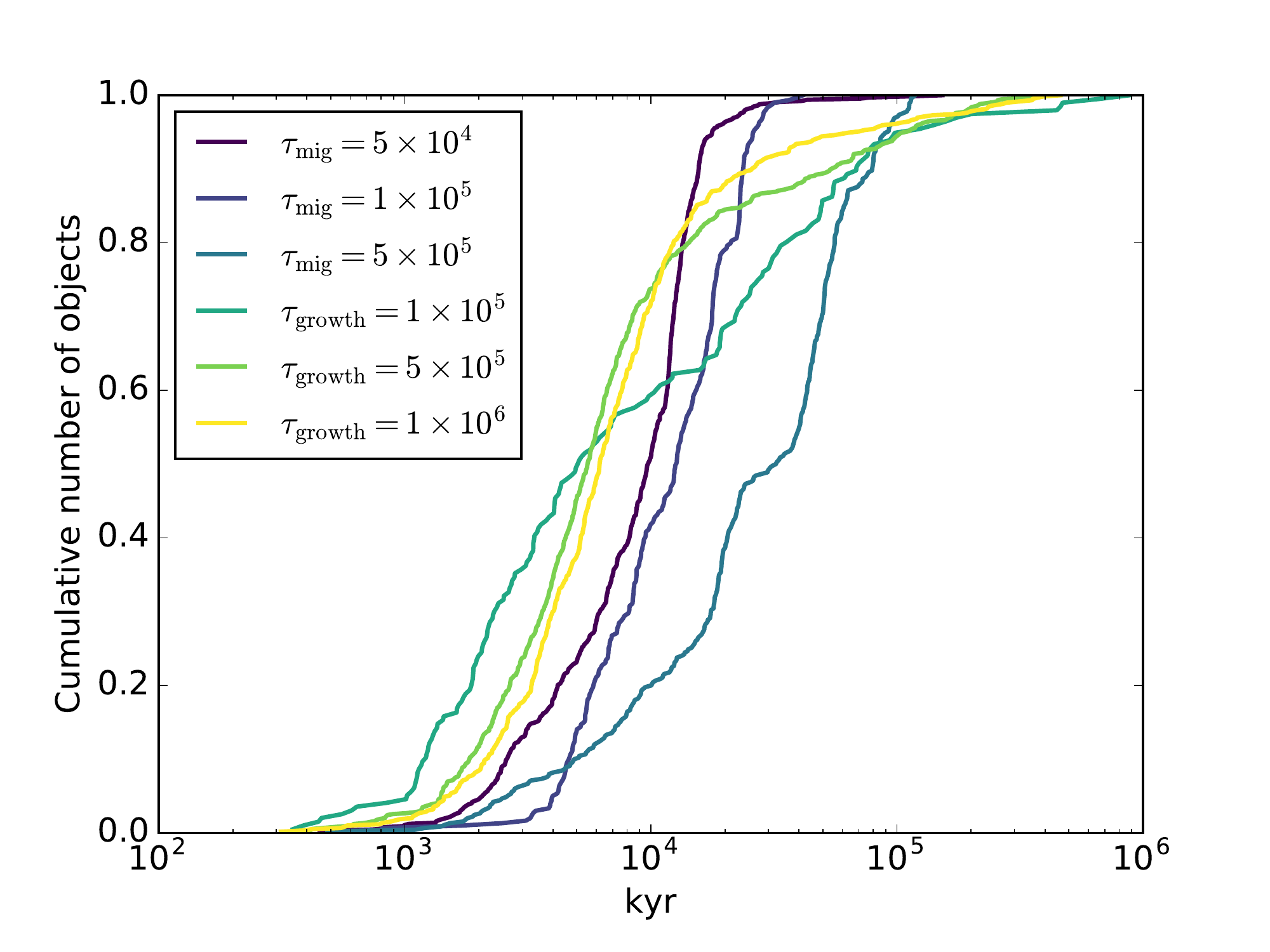}
\caption{Time evolution of the cumulative number of objects captured within Jupiter's CPD in Case 1 (formation of Saturn at the gap) and in Case 2 (migration of Saturn towards Jupiter), for different parameters investigated. In each scenario, the delivery of planetesimals to the circum-jovian disk spans a few $10^5$ years.}
\label{delivery}
\end{figure}

We now investigate the evolution of the planetesimals captured in orbits around Jupiter. For all cases considered, the planetesimals captured within the CPD have initially very eccentric and inclined orbits at large distances from Jupiter. Slightly less than half of the objects captured are actually found in retrograde orbits. Figure~\ref{initial} shows that the distributions of orbital parameters of the objects at the time of their capture are quite similar in the most favorable scenarios of cases 1 and 2. Similar trends were obtained by \citet{SO17} although they considered that planetesimals initially populate the close vicinity of Jupiter (i.e., the region inside of Jupiter's gap in our configuration) and no other massive object but Jupiter perturbed their orbits. It should be noted that the distribution of objects in Figure~\ref{initial} is not representative of the system at a particular time because the planetesimals were not all captured concurrently. 
The delivery of planetesimals actually spans $\sim$10$^4$--10$^5$ years depending on the adopted parameters (see Figure~\ref{delivery}).
     
\begin{figure}
\includegraphics[width=\linewidth]{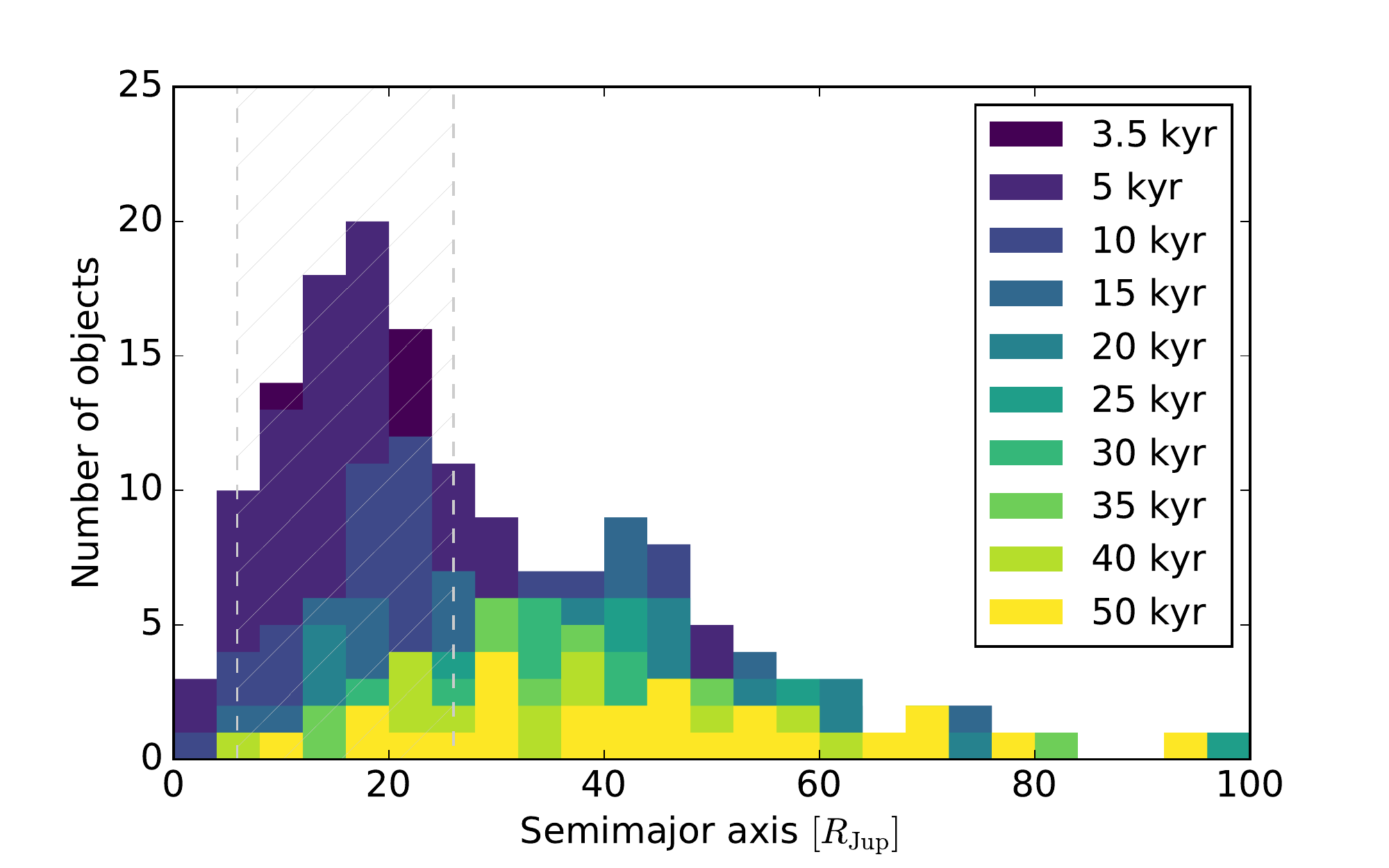}
\caption{Distribution of planetesimals at different epochs in Case 1 with $\tau_\mathrm{growth}=5\times10^5$ years. Each bin is 4 $R_\mathrm{Jup}$ wide. The hatched region indicates the present day extension of the Galilean system, with the inner and outer edges being the radial positions of Io ($\sim$5.9 $R_\mathrm{Jup}$) and Callisto ($\sim$26 $R_\mathrm{Jup}$).}
\label{evolution}
\end{figure}
  
To illustrate the subsequent evolution of the captured planetesimals, we conducted simulations centered on Jupiter as the only massive object and integrated the orbits of the planetesimals within the CPD for the most favorable scenario of Case 1. The simulation started at the time of capture of the first planetesimal and objects were subsequently added at their corresponding capture time as the simulation evolves. We also assumed a slightly subkeplerian velocity of the gas around Jupiter \citep[$v_\mathrm{orb}=(1-\eta)v_\mathrm{kep}$, where $\eta$ is a measure of the pressure support of the disk and we used $\eta=0.005$, typical for keplerian disks,][]{Jo14} to account for the potential loss of objects through inward drift due to gas drag. Figure~\ref{evolution} shows the distribution of the planetesimals as a function of their distance from Jupiter at different epochs of the CPD's evolution. Objects that are captured on initially retrograde orbits are rapidly lost to Jupiter due to gas drag. On the other hand, the planetesimals initially captured on prograde orbits with large eccentricities and inclinations rapidly circularize and pile up in the inner part of the CPD (c.f., the histogram drawing the distribution of captured objects after 5 kyr of evolution). The hatched region of Figure~\ref{evolution} illustrates the current extension of the Galilean system with the inner and outer radial boundaries being the position of Io and Callisto, respectively. Interestingly, the region where planetesimals pile-up matches well that where the Galilean satellites orbit.\\
  
After having rapidly reached a maximum at $\sim$5 kyr, the number of objects in the CPD slowly decreases as the planetesimals drift inward due to gas drag faster than the replenishment due to the capture of new objects. The decay is nevertheless slow compared to the orbital period of the objects which is $\sim$2 days at Io's orbit and $\sim$17 days at Callisto's orbit. The timescale of orbital decay due to gas drag can be estimated as $\tau_\mathrm{drag} = r\frac{\mathrm{d}t}{\mathrm{d}r}$ with $\frac{\mathrm{d}r}{\mathrm{d}t}=\frac{2\mathrm{St}}{1+\mathrm{St}^2}\eta v_\mathrm{kep}$ \citep[e.g.,][]{We77b}, with St the Stokes number of the planetesimal (i.e., the stopping time normalized by the keplerian frequency; see Appendix~\ref{extra} for an expression of the stopping time). Considering that St $\gg 1$, relevant for large planetesimals, the decay timescale can be expressed as :
\begin{equation}
  \tau_\mathrm{drag} \sim \frac{1}{2} \mathrm{St} \frac{T_\mathrm{orb}}{2\pi\eta} \sim 1.6\times10^7 \left(\frac{St}{10^6}\right) \left(\frac{0.005}{\eta}\right) T_\mathrm{orb}
\end{equation} 
In the above expression $T_\mathrm{orb}$ is the orbital period of the object. On the other hand, \citet{CW02} approximate a satellite's growth timescale as :
\begin{equation}
  \tau_\mathrm{acc} \sim 8\times10^6 \left(\frac{\rho_s}{2 \,\mathrm{g\,cm^{-3}}}\right) \left(\frac{R_\mathrm{sat}}{2500\, \mathrm{km}}\right) \left(\frac{1 \,\mathrm{g\,cm^{-2}}}{\Sigma_s}\right) \left(\frac{10}{F_g}\right) T_\mathrm{orb}.
\end{equation} 
In the latter expression, $\rho_s$ is the mass density of the satellite, $R_\mathrm{sat}$ its radius, $\Sigma_s$ is the surface density of solids within the CPD and $F_g = 1+(v_\mathrm{esc}/v_\mathrm{rel})^2$ is the gravitational focusing factor with $v_\mathrm{rel}$ the relative velocity between satellitesimals and $v_\mathrm{esc}$ their mutual escape velocity. Therefore, the collisional growth of the objects should have been efficient provided that the surface density of solids was at least of the order of 1 g$\,$cm$^{-2}$ which is a rather low value appropriate for starved-disk formation models.

\section{Discussion} 
\label{discussion}
  
\subsection{Implantation of planetesimals in the asteroid belt}
  
\begin{figure}
\includegraphics[width=\linewidth]{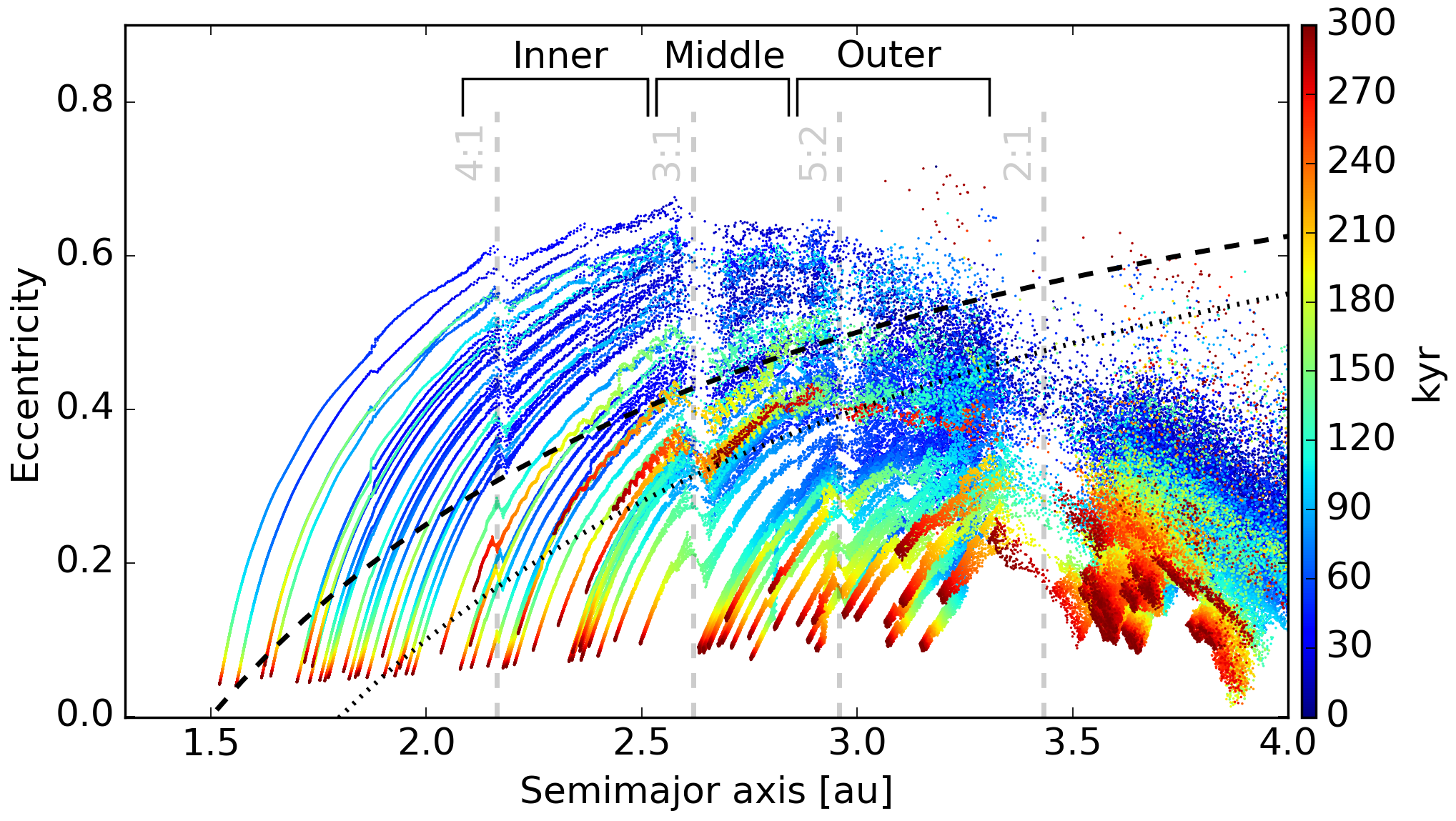}
\caption{Trajectories of planetesimals in the semimajor axis-eccentricity plane of the asteroid belt region. The colors of the dots give an indication of the time. The dotted and dashed lines mark the limit where the periapsis of the orbit is $q = 1.8$ AU (roughly the edge of the asteroid belt) and $q = 1.5$ AU (region of the terrestrial planets' embryos), respectively. The positions of major mean motion resonances with Jupiter are represented by the vertical dashed lines. These are the resonances that define today's asteroid belt regions, labelled Inner, Middle and Outer in the figure. The different regions are shifted inward as compared to the position of the MMRs because Jupiter is orbiting at $\sim$5.4 AU at the end of the simulation, consistently with models of later dynamical evolution of the outer Solar System.}
\label{MB}
\end{figure}

In Section~\ref{results}, we have shown that for both the formation of Saturn at the outer edge of Jupiter's gap and at further distances, planetesimals from the reservoir are redistributed accross the inner Solar System. Recently \citet{RI17} proposed that the redistribution of planetesimals by the gas giants is a natural outcome of their formation, providing an explanation for the delivery of water to the terrestrial planets and the presence of primitive C-type asteroids in the outer asteroid belt. The authors demonstrated that some planetesimals were always scattered inward of Jupiter's orbit regardless of the precise growth timescale or migration rates of Jupiter and Saturn in their simulations. However, the planetesimals were initially spread between 2--20 AU in their simulations , which is quite different from the distribution we consider in this work. Our results therefore support the findings of \citet{RI17}, showing them to be robust against more specific initial conditions and accounting for the fact that objects might be captured by Jupiter instead of being scattered inward of its orbit.\\

Figure~\ref{MB} represents the trajectories of planetesimals in the semimajor axis-eccentricity plane in the 1.2--4.0 AU region, along with important MMRs with Jupiter and the different regions of the main asteroid belt (inner, middle and outer belt). Our simulations show that planetesimals are not preferentially implanted in the outer region of the Main Belt, where the majority of C-type asteroids are found today. This result should nevertheless be considered with caution for several reasons. First, planetary embryos were not included in our simulations. The planetesimals ending up in the inner parts of the asteroid belt have trajectories that cross the embryos' region, marked by the dashed line in Figure~\ref{MB}. The final distribution of objects in the inner belt might be inaccurate due to the fact that the influence of embryos was not accounted for in this work. Second, the C-type spectral group embraces a great diversity of objects with potentially very different origins or formation times \citep[e.g.,][]{Ve17}. If the diversity among C-type asteroids indeed traces different origins, it is likely that the different populations were not implanted at the same time, or that some C-type asteroids have formed {\it in situ} so that not all of the objects from this group were actually implanted in the belt. Finally, we have not implemented the decay of the gas density due to the viscous evolution of the PPD and/or the photoevaporation of the disk. As the density decays, the damping due to gas drag is less efficient and planetesimals can reach more distant regions in the inner solar system \citep{RI17}. As we used a constant surface density, the planetesimals were implanted quite homogeneously from 1.5 to 3.5 AU in our simulations.
  
Instead of reasoning in terms of spectral types, \citet{Kr17} proposed that the observed dichotomy in the isotopic ratios of carbonaceous and non carbonaceous meteorites is due to the separation of the formation regions of the parent bodies of these meteorites by Jupiter's core. This way, the two reservoirs of objects could not mix and their isotopic differences were preserved. The authors were able to put new constraints on the formation timescales of the carbonaceous chondrites that would have formed beyond Jupiter's orbit. They showed that the formation of the parent bodies of the carbonaceous meteorites started $\sim$1 My after the condensation of the CAIs (Carbon and Aluminium rich Inclusions) and ended $\sim$4 My after CAIs, implying that the reservoir of carbonaceous material has been separated from that of non carbonaceous material for $\sim$3 My. These constraints can be matched in the framework of our scenario, suggesting that the formation of the parent bodies of carbonaceous chondrites was triggered by the end of the accretion of solid material onto Jupiter's core. From this moment, solids (in the form of pebbles) accumulated at the pressure perturbation induced by the forming planet and eventually collapsed into larger objects. Their injection in the inner solar system was then triggered by either the formation of Saturn's core or its migration in the vicinity of Jupiter. This would naturally account for the delay between the formation of the carbonaceous meteorites parent bodies and their mixing with the non carbonaceous meteorites parent bodies, which formed and remained inside of Jupiter's orbit  and were not included in our simulations.
  
\subsection{Accretion of the Galilean satellites}
  
In Section~\ref{evolution}, we have shown that the planetesimals rapidly pile-up in the region where the Galilean satellites are found today. This could provide suitable conditions for the rapid formation of satellite seeds in this region. The satellites would then fully accrete on longer timescales ($\sim$10$^5$ years, Figure~\ref{delivery}), limited by the capture of new objects by Jupiter and the slow orbital decay of the planetesimals that have been circularized on wider orbits. Such a scenario would be consistent with a partially differentiated Callisto \citep{BC08}.
  
In a previous study, \citet{RMV17} have shown that the compositional gradient among the Galilean satellites could be accounted for if they accreted from pebbles with sizes ranging 1--10$^2$ cm. This scenario, however, implies that the pebbles do not disintegrate as their water ice starts to sublimate when crossing the snowline and the migration of Europa was tied to the evolution of the snowline. In the present study, we used objects with a radius of 100 km as typical planetesimals, a choice motivated by the existence of such large primitive objects in the asteroid belt, pointing toward the existence of a planetesimal reservoir outside of Jupiter's orbit. At first sight, this seems contradictory with the pebble accretion scenario proposed by \citet{RMV17}. However the planetesimals captured within the CPD being initially on very excited, both prograde and retrogade orbits, violent impacts could have led to an intense grinding of the planetesimals. It is therefore plausible that a non negligible amount of material was found in objects with a size in the meter range and below and were subsequently efficiently accreted by the larger objects that did not suffer disruptive collisions. This would also be favorable to the formation of an only partially differentiated Callisto as noted by several authors \citep[e.g.,][]{LS82,BC08}. It should be also noted that, considering the dynamical state of the reservoir, disruptive collisions within the reservoir might have provided an important source of dust grains. \citet{Sh17} pointed out in their study the difficulty in growing large objects from dust grains within CPDs, hinting towards the existence of already large objects that would act as the seeds of the protosatellites. The Galilean satellites might well have grown through a combination of planetesimal and pebble accretion.
  
\subsection{Effect of the surface density of the CPD}
  
Our nominal set of simulations was performed using a CPD with a surface density that is approximately one order of magnitude higher than that of the gas-starved disk proposed by \citet{CW02,CW06}, with a peak surface density at $\sim$10$^4$ $\mathrm{g \, cm}^{-2}$. Such surface densities are still lower than that adopted in the minimum mass model \citep[surface density peak at $\sim$10$^6$ $\mathrm{g \, cm}^{-2}$,][]{ME03a,ME03b}. The disk profile we used is likely representative of the stage when Jupiter is still feeding from the surrounding nebula \citep[e.g.,][]{Fu17}. However, as the PPD's density is supposedly decaying and Jupiter's gap deepening over time, the surface density of the CPD would also decay, leading to a less efficient capture of planetesimals through gas drag. Therefore, the capture efficiencies may be lower than obtained here.
  
To investigate whether our results are realistic, we ran the Case 1 and Case 2 scenarios with the optimal parameters, namely $\tau_\mathrm{growth}= 5\times10^5$ years for Case 1 and $\tau_\mathrm{mig}=10^5$ years for Case 2, with a CPD profile identical to that of \citet{Sa10} (cf. Appendix~\ref{CPDmod}). These authors investigated the growth of the Galilean satellites with a semi-analytical model in the context of a slightly modified starved disk scenario. In both simulations, the CPD capture efficiencies dropped to $\sim$8\%. Such efficiencies are still in the range of values obtained by varying Saturn's growth or migration timescale.
  
In Section~\ref{CPD} we showed that planetesimals are delivered over a $\sim$10$^5$ years timescale. The capture of large planetesimals would therefore remain efficient if the CPD's surface density does not decay significantly during this timescale (i.e., Jupiter is still accreting gas from the PPD and/or the viscous evolution of the CPD is slow). A more subtle effect that has been ignored in the present study is that planetesimals with different sizes would have different capture efficiencies due to a more or less efficient gas drag braking within the circum-jovian disk. The evolution of the CPD's surface density would likely results in an evolution of the size distribution of captured objects which could affect the subsequent growth of the satellites. More detailed studies, including plausible planetesimals size distributions at the outer edge of Jupiter's gap and evolution of the circum-jovian disk, are needed to determine more realistic conditions of accretion of the Galilean satellites.

\subsection{Influence of Saturn's growth track}

Although we varied Saturn's growth timescale by an order of magnitude when investigating the dynamical evolution of planetesimals in Section~\ref{gap}, the use of equation~(\ref{Mdot}) always implies that the mass doubling timescale of the planet is shorter in the early phases of its growth. As demonstrated by \citet{SI08}, a growing planet generally experiences more close encounters with nearby planetesimals if its mass doubling timescale is shorter because the expansion of its Hill sphere is then fast compared to the gap opening timescale in the planetesimals' disk. If the growth of Saturn was initially slow enough, the protoplanet might have carved a gap in the planetesimal's distribution which would have prevented an efficient scattering and delivery of the planetesimals towards Jupiter. Hence, the use of equation~(\ref{Mdot}) might overestimate the ability of Saturn's core to scatter nearby planetesimals in the early phases of its growth. We note however that if Saturn's core had grown through pebble accretion, its mass doubling timescale would have indeed been shorter in the early phases of its growth \citep[due to the sublinear dependance of the pebbles accretion rate on the mass of the core,][]{LJ12} and certainly shorter than the gap opening timescale in the planetesimals' disk.

To assess the robustness of the redistribution of planetesimals against Saturn's growth track, we ran an additional simulation with a qualitatively different growth rate for Saturn. In this simulation, we let Saturn grow according to $M_\mathrm{sat}/\dot{M}_\mathrm{sat} = 10^6$ years, which yields a very slow initial growth (the mass of the protoplanet is $\sim$2$\,M_\oplus$ after $\sim$5$\times10^5$ years) and a rapid final assemblage of the planet. The capture efficiency within the CPD obtained was $\sim$11\%, which compares well with the results obtained using equation~(\ref{Mdot}). This is due to the fact that the opening of a gap within the planetesimals' disk by the growing core is prevented by nearby Jupiter which stirs the orbits of the objects in the reservoir, maintaining high eccentricities. It is therefore the combined influence of Jupiter and growing Saturn, and not uniquely Saturn's growth, which allows for an efficient redistribution of the planetesimals. The precise growth of Saturn hence has little effect on its ability to scatter nearby planetesimals. We note that an effect which might damp the eccentricities of the planetesimals and was not included in our simulations is collisions among the objects. Taking collisions into account would however require the assumption of an initial mass of the reservoir, considered as unknown in the present study. We leave such a different approach to the problem, and the investigation of the effects of collisions, to future work.

\subsection{Formation of Saturn's satellite system}
  
Saturn possesses a unique assemblage of regular satellites with a possible dual origin. The small satellites orbiting close to Saturn are thought to have formed from the spreading of ring material across the Roche radius while Titan and Iapetus could have formed via a mechanism similar to those invoked for the formation of the Galilean satellites \citep{CSC10,CC12,SC17}. 
 When the two gas giants were close together within the PPD (in their mutual 2:1 or 3:2 MMR), they would have opened a unique and large gap in the disk \citep{MC07,Pi14}. The solids would then be trapped outside of Saturn's orbit, at the outer edge of the common gap opened by Jupiter and Saturn. If enough material remained in the form of pebbles at this time in the PPD, a new reservoir of planetesimals could have built up there. Either the formation of the cores of Uranus and Neptune at the gap, or their migration towards Saturn, could have allowed the delivery of planetesimals from this new reservoir to Saturn's CPD to build Titan and Iapetus.
  
\subsection{Implications for the formation of extrasolar moons}
  
In this study, we have pointed out that the gap opened by a giant planet in a PPD efficiently isolates it from the main sources of solid material. In our proposed scenario, the delivery of solids to the giant planet's CPD results from the interaction of a massive object with a reservoir of planetesimals. From this perspective, it is to be expected that the formation of massive moons is not ubiquitous, especially in systems with single or isolated giant planets. Moreover, if a giant planet is orbiting close to its host star, its Hill sphere is reduced and the capture rate of planetesimals could be lowered due to larger orbital velocities, therefore acting against the formation of a massive satellite system. 
  
\section{Summary}\label{summary}
  
An important step in understanding the formation of the giant planet's satellite systems is to elucidate the origin and delivery mechanism of the solid material needed to build the moons. Here we attempted to revisit the origin and delivery of the building blocks of the Galilean satellites, based on our current understanding of giant planet formation. Our findings can be summarized as follows:

\begin{itemize}
\item[-] Based on studies by \citet{SO17} and \citet{Pa07,Zh12}, we concluded that the gap opened by Jupiter efficiently isolated the giant planet and its circumplanetary disk from sources of solid material such as pebbles or planetesimals. However, the accumulation of solids at the outer edge of the gap likely translated into a planetesimal reservoir there.

\item[-] The planetesimals' orbits were then excited by the formation of Saturn at Jupiter's gap or during its migration towards Jupiter.

\item[-] This triggered the redistribution of planetesimals from the reservoir to the circum-jovian disk and the inner Solar System, with a moderate dependency on the input parameters of our model such as the growth timescale of Saturn or its migration rate. Therefore, we find there exists a link between primitive asteroids of the Main Belt and the Galilean satellites, as they shared a common reservoir. This link could be a testable constraint of our scenario by future missions to the jovian system, such as the ESA Juice mission, as some isotopic correspondences (e.g., the D/H ratio in water) should exist between the satellites and the asteroids.

\item[-] We find that the planetesimals are initially captured on very eccentric, both prograde and retrograde orbits within the circum-jovian disk. The subsequent gas drag damping of the orbits  results in an accumulation of objects in the region where the Galilean satellites are found today.

\item[-] The decisive role of Saturn in the delivery of material to the jovian disk has severe implications for the occurence of massive moons around extrasolar giant planets. If our proposed scenario is correct, massive satellites would preferentially form around giant planets in multiple planet systems.
\end{itemize}

Finally, it appears difficult to disentangle the formation of Saturn at the outer edge of the gap opened by Jupiter from its formation further from Jupiter and subsequent migration considering only the implications for the formation of the Galilean moons. Both scenarios provide quite similar results, although we believe that our so-called Case 1 scenario provides a more consistent model for Saturn's formation. Additionnal constraints should come from more detailed studies of Saturn's growth and the implications of the different formation scenarios on its final composition. In the present study, we left aside some important issues such as the size distribution of planetesimals, the evolution of the circum-jovian disk or the accretion of the satellites. More detailed simulations are needed to assess realistic conditions for the accretion of Jupiter's massive moons.

\acknowledgements
The authors thank the anonymous referee for their comments that helped reinforce the present study. This work has been partly carried out thanks to the support of the A*MIDEX project (n\textsuperscript{o} ANR-11-IDEX-0001-02) funded by the ``Investissements d'Avenir'' French Government program, managed by the French National Research Agency (ANR). O.M. acknowledges support from CNES. JIL acknowledges support from the Juno project.

\appendix{ 
  
\section{Additional forces for planets and planetesimals}\label{extra}

Here we describe the effects of aerodynamic drag and eccentricity/semi-major axis damping that were included in our simulations. Following \citet{CN08}, we included the effects of eccentricity and semi-major axis damping of the planets due to interactions with the gas disk through the following acceleration term:

\begin{equation}
{\bf a}_\mathrm{mig} = -\frac{\bf v}{\tau_\mathrm{mig}},
\end{equation}
  
\begin{equation}
{\bf a}_\mathrm{e} = -2\frac{({\bf v \cdot r }){\bf r}}{r^2 \tau_\mathrm{e}}.
\end{equation}
  
\noindent In the above expressions, ${\bf v}$ is the velocity vector of the planet, ${\bf r}$ its position vector and $r$ the distance to the star. In the case of Saturn, the eccentricity damping timescale $\tau_\mathrm{e}$ was taken to be $0.01\, \tau_\mathrm{mig}$ \citep[e.g.,][]{LP02}. As we did not consider any radial migration of Jupiter, we always used an eccentricity damping timescale of $\tau_\mathrm{e}=5\times10^3$ years and no semi-major axis damping for this planet. These are simplified prescriptions that do not take into account the structure of the disk. However the purpose of this study is not to investigate the precise migration of the giant planets within the disk.
  
  We accounted for the aerodynamic drag effects on the planetesimals. This was implemented in a similar fashion as in \citet{RMV17} by adding the following acceleration term:
\begin{equation}
{\bf a}_\mathrm{drag} = - \frac{1}{t_s}({\bf v - v}_g).
\end{equation}
  
\noindent In the above expression, ${\bf v}_g$ is the velocity of the gas given by the hydrodynamic simulation when planetesimals are far from Jupiter. When planetesimals are at a distance of 150 $R_\mathrm{Jup}$ from Jupiter or closer, the gas velocity is found assuming a keplerian velocity around the giant planet to model the interaction with the CPD. The stopping time $t_s$ is computed using the following expression \citep{PM11,GIO14}: 

\begin{equation}
t_s = \left( \frac{\rho_g v_{th}}{\rho_s R_s} \mathrm{min}\left[ 1, \frac{3}{8} \frac{v_{rel}}{v_{th}} C_D(Re)\right] \right)^{-1}.
\end{equation}
  
\noindent In this expression, $R_s$ is the size of the planetesimal and $\rho_s = 1 \; \mathrm{g\,cm^{-3}}$ its density.
  The gas density $\rho_g$ is obtained by assuming hydrostatic equilibrium in the vertical direction with an aspect ratio of the disk $h = 0.05$ in the case of the PPD or it is given by the CPD prescription described in the next section when planetesimals are close to Jupiter. The gas thermal velocity is $v_{th} = \sqrt{8/\pi}c_g$, $c_g$ is the isothermal sound speed and $v_{rel}$ is the relative velocity between the gas and the planetesimal, either in the CPD or the PPD. The dimensionless drag coefficient $C_D$ is computed as a function of the Reynolds number $Re$ of the flow around the planetesimal \citep{PM11}:
  
\begin{equation}
C_D=\frac{24}{Re} (1+0.27Re)^{0.43}+0.47\left(1-e^{-0.04Re^{0.38}}\right),
\end{equation} 
  
\begin{equation}
Re = \frac{4 R_s v_{rel}}{c_g l_g}.
\end{equation}

\noindent The mean free path of the gas $l_g$ is taken from the prescription of \citet{SL00}.
  
\section{The CPD model} \label{CPDmod}
  
\begin{figure}
\includegraphics[scale=0.5]{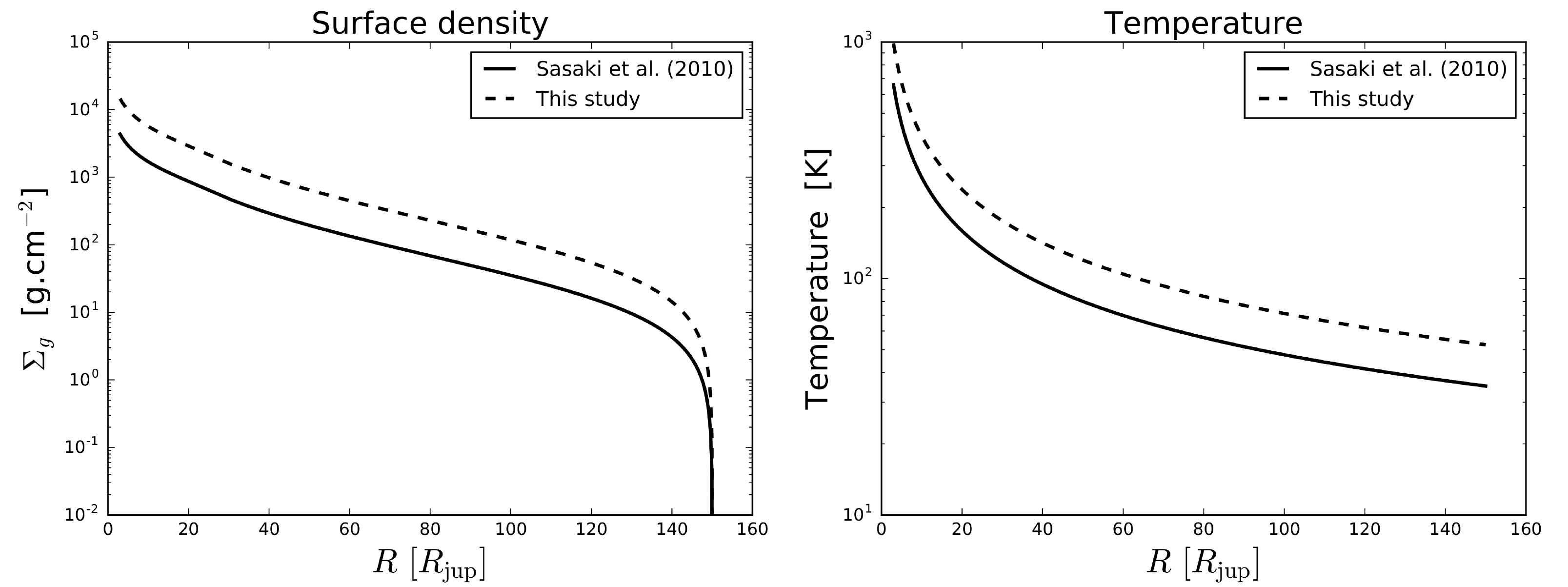}
\caption{{\it Left.} Comparison between the surface density profile used by \citet{Sa10} (solid line), obtained with $\dot{M}_p = 2\times10^{-7}\;M_\mathrm{Jup}\,$yr$^{-1}$, and the profile used in this study (dashed line) obtained with $\dot{M}_p = 1\times10^{-6}\;M_\mathrm{Jup}\,$yr$^{-1}$.{\it Right} Comparison between the temperature profiles assumed in this study (dashed line) and that assumed by \citet{Sa10}.}
\label{comparison}
\end{figure}
  
Our model is based on the simple prescription of \citet{Sa10}, which was constructed from the gas-starved model of \citet{CW02}. The surface density of the disk is found by considering an equilibrium between the mass accreted from the PPD to the CPD and the mass accretion rate onto Jupiter $\dot{M}_p$. The gas accreted from the PPD is considered to fall uniformly from the inner edge of the disk out to the centrifugal radius $R_c$ which is set at 26 $R_\mathrm{Jup}$. This gives the following expression for the surface density of the CPD \citep[e.g.,][]{CW02}:

\begin{equation}
\Sigma_g (r) = \frac{\dot{M}_p}{3\pi\nu(r)} \left \lbrace \begin{array}{ll}
1 - \frac{4}{5} \sqrt{\frac{R_c}{R_d}} - \frac{1}{5} \left(\frac{r}{R_c}\right)^2 & \mathrm{for} \; r\leqslant R_c\\
\\
\frac{4}{5} \sqrt{\frac{R_c}{r}} - \frac{4}{5} \sqrt{\frac{R_c}{R_d}} & \mathrm{for} \; r > R_c,\\
\end{array} \right.
\end{equation}

\noindent Here, $R_d = 150 \, R_\mathrm{Jup}$ is the outer radius of the disk and $\nu$ is the turbulent viscosity parameterized with the $\alpha$ equivalent turbulence $\nu = \alpha H^2_g \Omega_K$ \citep{SS73} and $\alpha = 10^{-3}$, where $H_g = c_s/\Omega_K$ is the disk scale height, $c_s=\sqrt{R_g T_d/\mu}$ is the gas isothermal sound speed with $R_g$ the ideal gas constant, $\mu = 2.4$ g$\,$mol$^{-1}$ the molecular weight of the gas, $T_d$ the temperature of the disk and $\Omega_K$ the keplerian frequency. The temperature profile of the disk is given by a balance between viscous dissipation and energy radiated away. Using the simplifications introduced by \citet{Sa10}, the temperature profile can be expressed as a function of the mass accretion rate :

\begin{equation}
T_d \simeq 225 \left( \frac{r}{10 \, R_\mathrm{Jup}} \right)^{-3/4} \,  \left( \frac{\dot{M}_p}{10^{-7} \, M_\mathrm{Jup}\,\mathrm{yr}^{-1}} \right)^{1/4} \; \mathrm{K} . 
\end{equation} 

\noindent More details can be found in the work by \citet{Sa10} \citep[see also,][]{RMV17}. Both the surface density and the temperature are therefore determined by the mass accretion rate onto Jupiter $\dot{M}_p$. The nominal value of the accretion rate at the time of Galilean satellites formation assumed by \citet{Sa10} was $2\times10^{-7}\;M_\mathrm{jup} \, $yr$^{-1}$. In this work, we have assumed a mass accretion rate onto Jupiter $\dot{M}_p=10^{-6}\;M_\mathrm{jup} \, $yr$^{-1}$, resulting in a denser and hotter disk. We used this parameter because a denser disk allows a higher capture rate and this is also in line with the results of 3D hydrodynamic simulations where denser disks are found \citep[see e.g.][]{TOM12}. The results of hydrodynamic simulations should be considered with caution and could be more representative of the very early phase of the CPD. Nevertheless, \citet{Fu17} show that the turbulence of the disk should be weak due to an inefficient ionization of the gas. A lower turbulence results in a denser CPD for a given accretion rate. Therefore, it seems likely that the disk had a surface density slightly higher than that advocated in the starved model of \citet{CW02}. Figure~\ref{comparison} shows a comparison between the surface density and temperature profiles we used and those of the study by \citet{Sa10}.

\section{Capture of planetesimals} \label{capture}

\begin{figure}
  \centering
  \includegraphics[width = .8\linewidth]{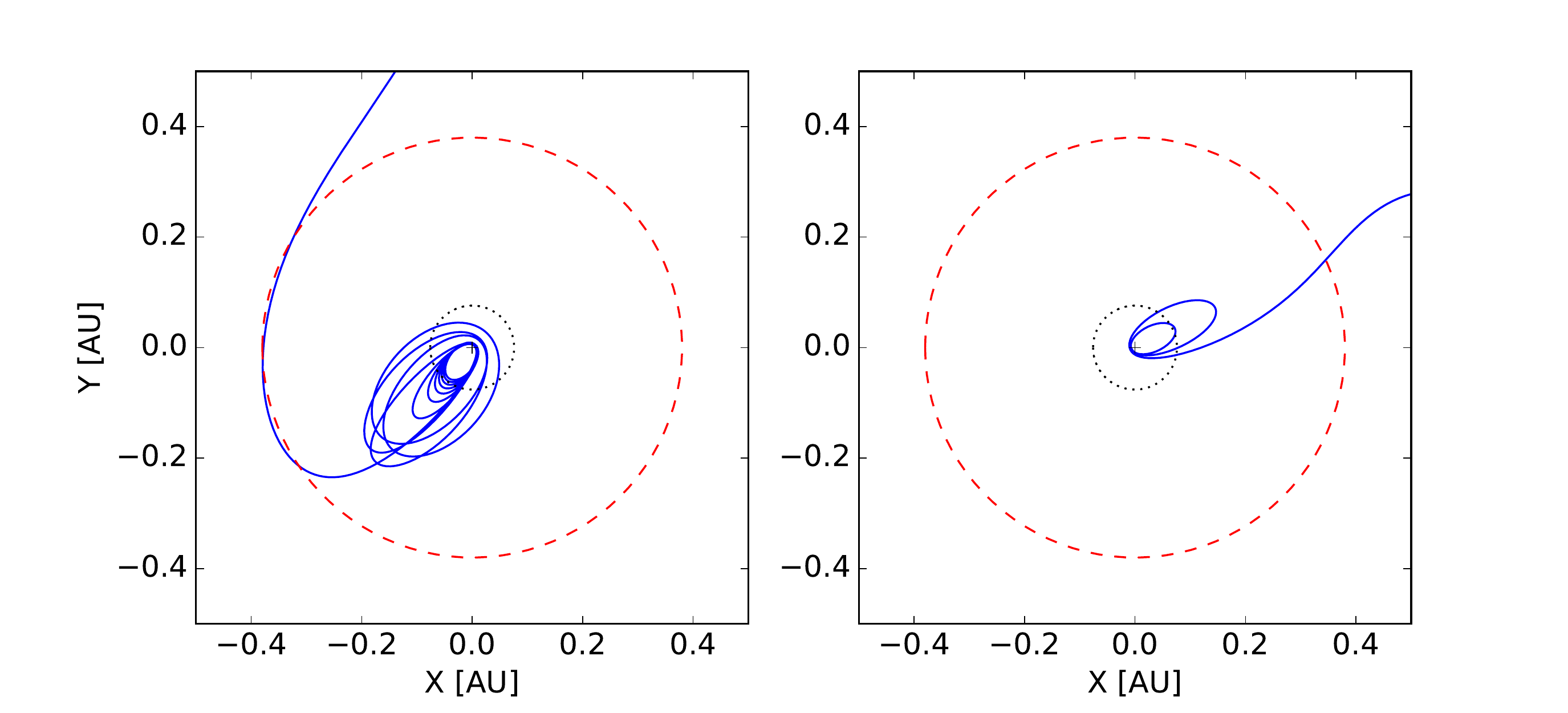}
  \caption{Orbits of captured planetesimals in a cartesian plane centered on Jupiter. The dashed red circle is Jupiter's Hill sphere whereas the dotted black circle shows the extension of the CPD. {\it Left :} Orbit of a planetesimal captured in the prograde direction with respect to Jupiter. {\it Right :} Orbit of a planetesimal captured in the retrograde direction.}
  \label{orbits}
\end{figure}

\begin{figure}
  \centering
  \includegraphics[width = .8\linewidth]{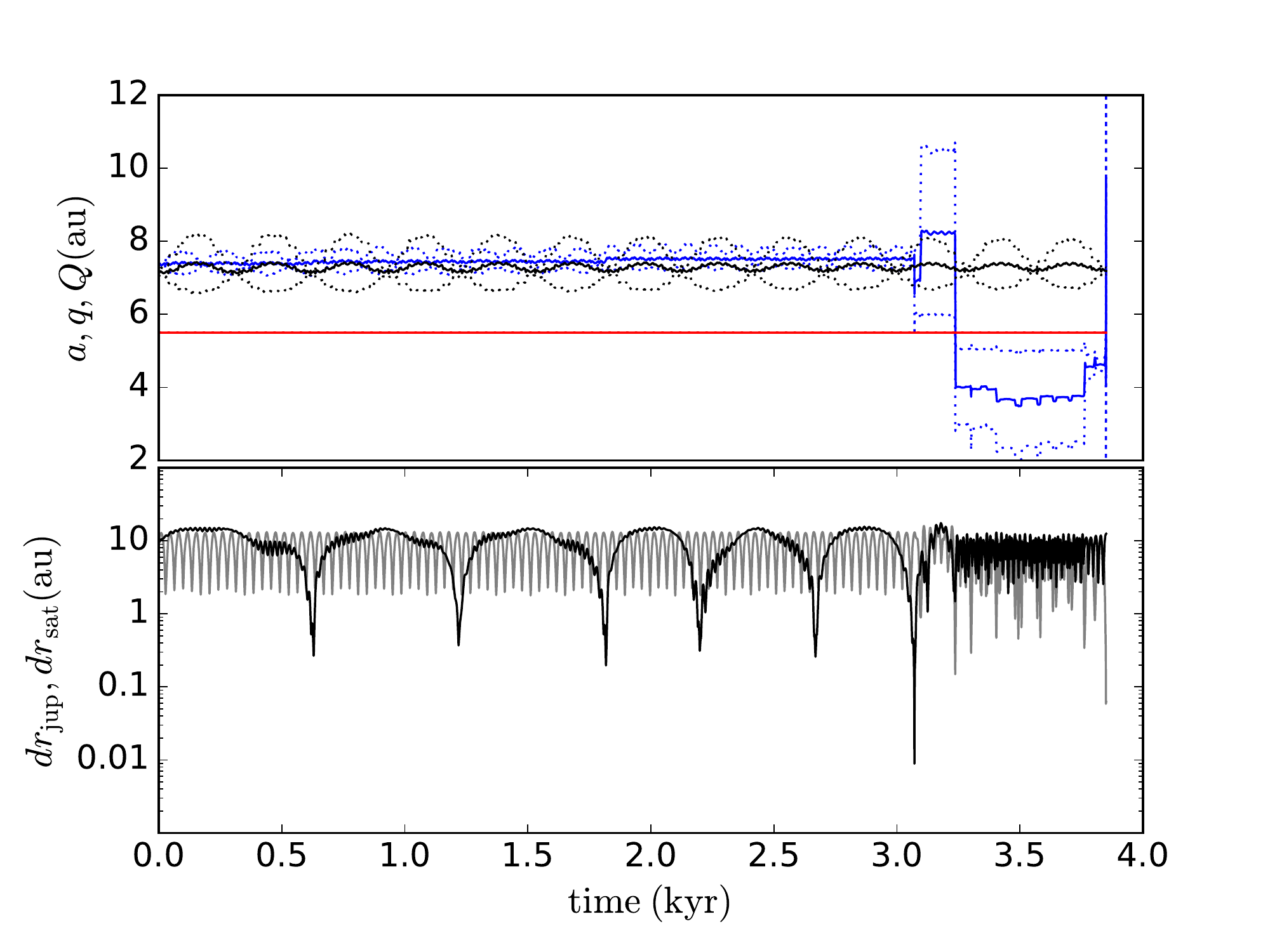}
  \caption{{\it Top :} Evolution of the semimajor axis $(a)$, perihelion distance $(q)$ and aphelion distance $(Q)$ of Jupiter (red), Saturn's core (black) and a planetesimal (blue) finally captured by Jupiter. {\it Bottom :} Evolution of the radial distance of the planetesimal relative to Jupiter (gray) and Saturn (black).}
  \label{orb_evol}
\end{figure}

We present here further details about the capture of planetesimals within the circum-jovian disk. More detailed investigations on the capture process can be found in the studies by \citet{Fu13}, \citet{SOF16} and \citet{SO17}. Gas drag is not efficient enough to allow for the direct capture within the circum-jovian disk of large planetesimals, such as those investigated in the present study, over a single passage through the CPD. Therefore, planetesimals experience a phase where they are captured on large orbits with respect to the extension of the CPD. During this phase, they cross the circum-jovian disk multiple times and their orbit gradually shrinks. Because the drag experienced by a planetesimal having a retrograde orbit with respect to Jupiter is much more efficient than that experienced in the case of prograde orbits (due to the lower relative velocity between the gas and the planetesimal in the latter case), planetesimals on retrograde orbits are more rapidly captured inside the CPD. They are however subsequently rapidly lost to Jupiter due to their fast orbital decay (see Sec.~\ref{CPD}). This is illustrated in Figure~\ref{orbits} where the orbits of planetesimals captured in the prograde (left panel) and retrograde (right panel) directions are showed. The orbits of these objects were integrated until they were found on bound orbits with a semimajor axis with respect to Jupiter that is smaller than 0.1 $R_\mathrm{Hill}$. The orbits were taken from a Case 1 simulation with Saturn growing on a $5.5\times 10^5$ year timescale. The planetesimal captured in the prograde direction clearly experienced many more CPD-crossing orbits before reaching our capture threshold than its sibling captured on a retrograde orbit. We note that the capture of large planetesimals, although dependant on their initial energy, generally requires that the object approach Jupiter at a distance $\lesssim 10^{-2} \, R_\mathrm{Hill}$ for the CPD's parameters adopted here.
To test the sensitivity of the capture efficiencies presented in the main text on the capture threshold imposed, we ran a full Case 1 simulation with a 0.1 $R_\mathrm{Hill}$ capture threshold. We obtained a capture efficiency of 14.6\%, in very good agreement with the results obtained using the less restrictive threshold presented in Section~\ref{gap}.

 Figure~\ref{orb_evol} shows an exemple of the heliocentric orbital evolution of a planetesimal before it is captured within the jovian CPD. The top panel shows the evolution of the semimajor axis (solid lines), perihelion and aphelion distances (dotted lines) of Jupiter (red), Saturn's core (black) and the planetesimal (blue). The bottom panel shows the corresponding evolution of the radial distance of the planetesimal relative to Jupiter (gray line) and Saturn's core (black line). Initially, the semimajor axis and the eccentricity of both Saturn's core and the planetesimal oscillate due to their proximity with the outer 3:2 MMR with Jupiter located at $\sim$7.2 AU. The planetesimal experiences a close encounter with Saturn's core after $\sim$3.05 kyr, which can be identified in the bottom panel of Figure~\ref{orb_evol}. This interaction yields an abrupt change of the semimajor axis of the planetesimal, from $\sim$7.5 to $\sim$6.9 AU, and an increase of the eccentricity, originally varying around a value of $\sim$0.03, up to a value of $\sim$0.13. This event triggers a more chaotic evolution of the planetesimal which interacts with Jupiter several times, further increasing its eccentricity to values close to 0.4 after it is scattered inward of Jupiter's orbit at 3.24 kyr. Interestingly, the planetesimal experiences two encounters with Jupiter soon before it is captured, at 3.76 and 3.81 kyr, both bringing its semimajor axis closer to that of Jupiter and reducing its eccentricity down to a value of $\sim$0.04. Due to the chaotic evolution of the planetesimals before their capture, a typical evolution is not easy to define but we find that captured planetesimals generally experience a close encounter with Saturn's core, triggering a chaotic phase of evolution during which their eccentricity is high and they interact several times with Jupiter. We find that the eccentricity of a planetesimal is often reduced following a close encounter with Jupiter right before the object is captured and is generally $\lesssim$ 0.2 then.


\begin{thebibliography}{}
    \bibitem[Barr \& Canup(2008)]{BC08} Barr, A.~C., \& Canup, R.~M.\ 2008, \icarus, 198, 163
    \bibitem[B{\'e}thune et al.(2016)]{BLF16} B{\'e}thune, W., Lesur, G., \& Ferreira, J.\ 2016, \aap, 589, A87  
    \bibitem[Birnstiel et al.(2011)]{BOD11} Birnstiel, T., Ormel, C.~W., \& Dullemond, C.~P.\ 2011, \aap, 525, A11 
    \bibitem[Birnstiel et al.(2012)]{BKE12} Birnstiel, T., Klahr, H., \& Ercolano, B.\ 2012, \aap, 539, A148 
    \bibitem[Bitsch et al.(2015a)]{Bi15} Bitsch, B., Johansen, A., Lambrechts, M., \& Morbidelli, A.\ 2015, \aap, 575, A28 
    \bibitem[Bitsch et al.(2015b)]{BLJ15} Bitsch, B., Lambrechts, M., \& Johansen, A.\ 2015, \aap, 582, A112 
    \bibitem[Canup \& Ward(2002)]{CW02} Canup, R.~M., \& Ward, W.~R.\ 2002, \aj, 124, 3404
    \bibitem[Canup \& Ward(2006)]{CW06} Canup, R.~M., \& Ward, W.~R.\ 2006, \nat, 441, 834 
    \bibitem[Canup \& Ward(2009)]{CW09} Canup, R.~M., \& Ward, W.~R.\ 2009, ~University of Arizona Press, Tucson, Ariz., 59 
    \bibitem[Carrera et al.(2015)]{CJD15} Carrera, D., Johansen, A., \& Davies, M.~B.\ 2015, \aap, 579, A43 
    \bibitem[Carrera et al.(2017)]{Ca17} Carrera, D., Gorti, U., Johansen, A., \& Davies, M.~B.\ 2017, \apj, 839, 16
    \bibitem[Charnoz et al.(2010)]{CSC10} Charnoz, S., Salmon, J., \& Crida, A.\ 2010, \nat, 465, 752 
    \bibitem[Cloutier et al.(2015)]{CTV15} Cloutier, R., Tamayo, D., \& Valencia, D.\ 2015, \apj, 813, 8 
    \bibitem[Coradini et al.(1995)]{Co95} Coradini, A., Federico, C., Forni, O., \& Magni, G.\ 1995, Surveys in Geophysics, 16, 533
    \bibitem[Cresswell \& Nelson(2008)]{CN08} Cresswell, P., \& Nelson, R.~P.\ 2008, \aap, 482, 677 
    \bibitem[Crida et al.(2006)]{CMM06} Crida, A., Morbidelli, A., \& Masset, F.\ 2006, \icarus, 181, 587  
    \bibitem[Crida \& Charnoz(2012)]{CC12} Crida, A., \& Charnoz, S.\ 2012, Science, 338, 1196
    \bibitem[Crida \& Bitsch(2017)]{CB17} Crida, A., \& Bitsch, B.\ 2017, \icarus, 285, 145 
    \bibitem[D'Angelo \& Podolak(2015)]{DaP15} D'Angelo, G., \& Podolak, M.\ 2015, \apj, 806, 203
    \bibitem[Deienno et al.(2014)]{De14} Deienno, R., Nesvorn{\'y}, D., Vokrouhlick{\'y}, D., \& Yokoyama, T.\ 2014, \aj, 148, 25 
    \bibitem[Deienno et al.(2017)]{De17} Deienno, R., Morbidelli, A., Gomes, R.~S., \& Nesvorn{\'y}, D.\ 2017, \aj, 153, 153 
    \bibitem[Dipierro et al.(2016)]{Di16} Dipierro, G., Laibe, G., Price, D.~J., \& Lodato, G.\ 2016, \mnras, 459, L1  
    \bibitem[Dipierro \& Laibe(2017)]{DL17} Dipierro, G., \& Laibe, G.\ 2017, \mnras, 469, 1932
    \bibitem[Dobos et al.(2017)]{DHT17} Dobos, V., Heller, R., \& Turner, E.~L.\ 2017, \aap, 601, A91  
    \bibitem[Dr{\c a}{\.z}kowska et al.(2016)]{DAM16} Dr{\c a}{\.z}kowska, J., Alibert, Y., \& Moore, B.\ 2016, \aap, 594, A105 
    \bibitem[Estrada \& Mosqueira(2006)]{EM06} Estrada, P.~R., \& Mosqueira, I.\ 2006, \icarus, 181, 486  
    \bibitem[Estrada et al.(2009)]{EM09} Estrada, P. R., Mosqueira, I., Lissauer, J. J., D’Angelo, G., \& Cruikshank, D. P. \ 2009, Uni-University of Arizona Press, Tucson, Ariz., 27
    \bibitem[Fujii et al.(2017)]{Fu17} Fujii, Y.~I., Kobayashi, H., Takahashi, S.~Z., \& Gressel, O.\ 2017, \aj, 153, 194 
    \bibitem[Fujita et al.(2013)]{Fu13} Fujita, T., Ohtsuki, K., Tanigawa, T., \& Suetsugu, R.\ 2013, \aj, 146, 140 
    \bibitem[Gomes et al.(2005)]{Go05} Gomes, R., Levison, H.~F., Tsiganis, K., \& Morbidelli, A.\ 2005, \nat, 435, 466 
    \bibitem[Gonzalez et al.(2015)]{Go15} Gonzalez, J.-F., Laibe, G., Maddison, S.~T., Pinte, C., \& M{\'e}nard, F.\ 2015, \planss, 116, 48 
    \bibitem[Guillot et al.(2014)]{GIO14} Guillot, T., Ida, S., \& Ormel, C.~W.\ 2014, \aap, 572, A72  
    \bibitem[Hansen(2009)]{Ha09} Hansen, B.~M.~S.\ 2009, \apj, 703, 1131 
    \bibitem[Hayashi(1981)]{Ha81} Hayashi, C.\ 1981, Progress of Theoretical Physics Supplement, 70, 35
    \bibitem[Heller et al.(2014)]{He14} Heller, R., Williams, D., Kipping, D., et al.\ 2014, Astrobiology, 14, 798 
    \bibitem[Heller \& Pudritz(2015)]{HP15} Heller, R., \& Pudritz, R.\ 2015, \apj, 806, 181  
    \bibitem[Izidoro et al.(2016)]{Iz16} Izidoro, A., Raymond, S.~N., Pierens, A., et al.\ 2016, \apj, 833, 40 
    \bibitem[Johansen \& Lacerda(2010)]{JL10} Johansen, A., \& Lacerda, P.\ 2010, \mnras, 404, 475 
    \bibitem[Johansen et al.(2014)]{Jo14} Johansen, A., Blum, J., Tanaka, H., et al.\ 2014, Protostars and Planets VI, 547 
    \bibitem[Johansen et al.(2015)]{Jo15} Johansen, A., Mac Low, M.-M., Lacerda, P., \& Bizzarro, M.\ 2015, Science Advances, 1, 1500109 
    \bibitem[Kobayashi et al.(2012)]{KOI12} Kobayashi, H., Ormel, C.~W., \& Ida, S.\ 2012, \apj, 756, 70 
    \bibitem[Kipping et al.(2015)]{Ki15} Kipping, D.~M., Schmitt, A.~R., Huang, X., et al.\ 2015, \apj, 813, 14 
    \bibitem[Krasinsky et al.(2002)]{Kr02} Krasinsky, G.~A., Pitjeva, E.~V., Vasilyev, M.~V., \& Yagudina, E.~I.\ 2002, \icarus, 158, 98 
    \bibitem[Kruijer et al.(2017)]{Kr17} Kruijer, T.~S., Burkhardt, C., Budde, G., \& Kleine, T.\ 2017, Proceedings of the National Academy of Science, 114, 6712 
    \bibitem[Lambrechts \& Johansen(2012)]{LJ12} Lambrechts, M., \& Johansen, A.\ 2012, \aap, 544, A32
    \bibitem[Lambrechts \& Johansen(2014)]{LJ14} Lambrechts, M., \& Johansen, A.\ 2014, \aap, 572, A107
    \bibitem[Lambrechts et al.(2014)]{LJM14} Lambrechts, M., Johansen, A., \& Morbidelli, A.\ 2014, \aap, 572, A35
    \bibitem[Lee \& Peale(2002)]{LP02} Lee, M.~H., \& Peale, S.~J.\ 2002, \apj, 567, 596 
    \bibitem[Levison et al.(2009)]{Le09} Levison, H.~F., Bottke, W.~F., Gounelle, M., et al.\ 2009, \nat, 460, 364
    \bibitem[Levison et al.(2010)]{LTD10} Levison, H.~F., Thommes, E., \& Duncan, M.~J.\ 2010, \aj, 139, 1297  
    \bibitem[Levison et al.(2015)]{LKD15} Levison, H.~F., Kretke, K.~A., \& Duncan, M.~J.\ 2015, \nat, 524, 322 
    \bibitem[Lin \& Papaloizou(1986)]{LP86} Lin, D.~N.~C., \& Papaloizou, J.\ 1986, \apj, 309, 846 
    \bibitem[Lunine \& Stevenson(1982)]{LS82} Lunine, J.~I., \& Stevenson, D.~J.\ 1982, \icarus, 52, 14 
    \bibitem[Machida et al.(2008)]{Ma08} Machida, M.~N., Kokubo, E., Inutsuka, S.-i., \& Matsumoto, T.\ 2008, \apj, 685, 1220-1236 
    \bibitem[Masset(2000)]{Ma00} Masset, F.\ 2000, \aaps, 141, 165 
    \bibitem[Morbidelli et al.(2005)]{Mo05} Morbidelli, A., Levison, H.~F., Tsiganis, K., \& Gomes, R.\ 2005, \nat, 435, 462 
    \bibitem[Morbidelli \& Crida(2007)]{MC07} Morbidelli, A., \& Crida, A.\ 2007, \icarus, 191, 158
    \bibitem[Morbidelli et al.(2014)]{Mo14} Morbidelli, A., Szul{\'a}gyi, J., Crida, A., et al.\ 2014, \icarus, 232, 266 
    \bibitem[Morbidelli et al.(2015)]{Mo15} Morbidelli, A., Walsh, K.~J., O'Brien, D.~P., Minton, D.~A., \& Bottke, W.~F.\ 2015, Asteroids IV, 493 
    \bibitem[Morbidelli \& Raymond(2016)]{MR16} Morbidelli, A., \& Raymond, S.~N.\ 2016, Journal of Geophysical Research (Planets), 121, 1962 
    \bibitem[Mosqueira \& Estrada(2003a)]{ME03a} Mosqueira, I., \& Estrada, P.~R.\ 2003, Icarus, 163, 198
    \bibitem[Mosqueira \& Estrada(2003b)]{ME03b} Mosqueira, I., \& Estrada, P.~R.\ 2003, Icarus, 163, 232
    \bibitem[Mosqueira et al.(2010)]{MEC10} Mosqueira, I., Estrada, P.~R., \& Charnoz, S.\ 2010, \icarus, 207, 448 
    \bibitem[Mousis \& Gautier(2004)]{MG04} Mousis, O., \& Gautier, D.\ 2004, \planss, 52, 361 
    \bibitem[Ormel \& Klahr(2010)]{OK10} Ormel, C.~W., \& Klahr, H.~H.\ 2010, \aap, 520, A43 
    \bibitem[Paardekooper \& Mellema(2006)]{PM06} Paardekooper, S.-J., \& Mellema, G.\ 2006, \aap, 453, 1129 
    \bibitem[Paardekooper(2007)]{Pa07} Paardekooper, S.-J.\ 2007, \aap, 462, 355
    \bibitem[Peale \& Canup(2015)]{PC15} Peale, S.~J., \& Canup, R.~M. \ 2015, Treatise on Geophysics (Second Edition), 559-604 
    \bibitem[Perets \& Murray-Clay(2011)]{PM11} Perets, H.~B., \& Murray-Clay, R.~A.\ 2011, \apj, 733, 56 
    \bibitem[Pierens et al.(2014)]{Pi14} Pierens, A., Raymond, S.~N., Nesvorny, D., \& Morbidelli, A.\ 2014, \apjl, 795, L11 
    \bibitem[Pollack et al.(1996)]{Po96} Pollack, J.~B., Hubickyj, O., Bodenheimer, P., et al.\ 1996, \icarus, 124, 62 
    \bibitem[Raymond \& Izidoro(2017)]{RI17} Raymond, S.~N., \& Izidoro, A.\ 2017, \icarus, 297, 134 
    \bibitem[Rein \& Spiegel(2015)]{RS15} Rein, H., \& Spiegel, D.~S.\ 2015, \mnras, 446, 1424 
    \bibitem[Rein \& Tamayo(2015)]{RT15} Rein, H., \& Tamayo, D.\ 2015, \mnras, 452, 376 
    \bibitem[Ronnet et al.(2017)]{RMV17} Ronnet, T., Mousis, O., \& Vernazza, P.\ 2017, \apj, 845, 92 
    \bibitem[Salmon \& Canup(2017)]{SC17} Salmon, J., \& Canup, R.~M.\ 2017, \apj, 836, 109 
    \bibitem[Sasaki et al.(2010)]{Sa10} Sasaki, T., Stewart, G.~R., \& Ida, S.\ 2010, \apj, 714, 1052
    \bibitem[Schoonenberg \& Ormel(2017)]{ScO17} Schoonenberg, D., \& Ormel, C.~W.\ 2017, \aap, 602, A21
    \bibitem[Shakura \& Sunyaev(1973)]{SS73} Shakura, N.~I., \& Sunyaev, R.~A.\ 1973, \aap, 24, 337
    \bibitem[Shibaike et al.(2017)]{Sh17} Shibaike, Y., Okuzumi, S., Sasaki, T., \& Ida, S.\ 2017, \apj, 846, 81
    \bibitem[Shiraishi \& Ida(2008)]{SI08} Shiraishi, M., \& Ida, S.\ 2008, \apj, 684, 1416-1426 
    \bibitem[Suetsugu et al.(2016)]{SOF16} Suetsugu, R., Ohtsuki, K., \& Fujita, T.\ 2016, \aj, 151, 140   
    \bibitem[Suetsugu \& Ohtsuki(2017)]{SO17} Suetsugu, R., \& Ohtsuki, K.\ 2017, \apj, 839, 66 
    \bibitem[Supulver \& Lin(2000)]{SL00} Supulver, K.~D., \& Lin, D.~N.~C.\ 2000, Icarus, 146, 525  
    \bibitem[Szul{\'a}gyi et al.(2014)]{Sz14} Szul{\'a}gyi, J., Morbidelli, A., Crida, A., \& Masset, F.\ 2014, \apj, 782, 65 
    \bibitem[Tanigawa et al.(2012)]{TOM12} Tanigawa, T., Ohtsuki, K., \& Machida, M.~N.\ 2012, \apj, 747, 47
    \bibitem[Teachey et al.(2017)]{TKS17} Teachey, A., Kipping, D.~M., \& Schmitt, A.~R.\ 2017, arXiv:1707.08563 
    \bibitem[Vernazza et al.(2017)]{Ve17} Vernazza, P., Castillo-Rogez, J., Beck, P., et al.\ 2017, \aj, 153, 72 
    \bibitem[Vokrouhlick{\'y} et al.(2016)]{VBN16} Vokrouhlick{\'y}, D., Bottke, W.~F., \& Nesvorn{\'y}, D.\ 2016, \aj, 152, 39 
    \bibitem[Walsh et al.(2011)]{Wa11} Walsh, K.~J., Morbidelli, A., Raymond, S.~N., O'Brien, D.~P., \& Mandell, A.~M.\ 2011, \nat, 475, 206 
    \bibitem[Weidenschilling(1977a)]{We77a} Weidenschilling, S.~J.\ 1977, \apss, 51, 153 
    \bibitem[Weidenschilling(1977b)]{We77b} Weidenschilling, S.~J.\ 1977, \mnras, 180, 57  
    \bibitem[Zhu et al.(2012)]{Zh12} Zhu, Z., Nelson, R.~P., Dong, R., Espaillat, C., \& Hartmann, L.\ 2012, \apj, 755, 6 
  \end{thebibliography}
\end{document}